\documentclass[letterpaper,11pt,aps,prd,amsmath,amssymb,floatfix,nofootinbib,reprint,superscriptaddress]{revtex4-1}
\usepackage{latexsym,bm}
\usepackage{graphicx}
\usepackage[utf8]{inputenc}
\usepackage{slashed}
\usepackage{amssymb}
\usepackage{amsmath}
\usepackage[dvipsnames]{xcolor}
\usepackage[colorlinks=true,allcolors=BlueViolet]{hyperref}
\usepackage{color} 
\usepackage{epsfig}
\usepackage{multirow}%
\usepackage{epstopdf}
\usepackage{tabularx}

\newcommand{\be}{\begin{equation}}
\newcommand{\ee}{\end{equation}}
\newcommand{\ba}{\begin{array}{c}} 
\newcommand{\ea}{\end{array}}
\newcommand{\bea}{\begin{eqnarray}} 
\newcommand{\eea}{\end{eqnarray}}

\begin{document}
\frenchspacing

\title{\LARGE Heavy-to-light scalar form factors from  Muskhelishvili--Omn\`es dispersion relations}

\author{D.-L.~Yao}
\email{deliang.yao@ific.uv.es}
\affiliation{Instituto de F\'{\i}sica Corpuscular (centro mixto CSIC-UV), Institutos de Investigaci\'on de Paterna,
Apartado 22085, 46071, Valencia, Spain}

\author{P.~Fernandez-Soler}
\email{pedro.fernandez@ific.uv.es}
\affiliation{Instituto de F\'{\i}sica Corpuscular (centro mixto CSIC-UV), Institutos de Investigaci\'on de Paterna,
Apartado 22085, 46071, Valencia, Spain}

\author{M.~Albaladejo}
\email{albaladejo@um.es}
\affiliation{Departamento de F\'{\i}sica, Universidad de Murcia, E-30071 Murcia, Spain}

\author{F.-K.~Guo}
\email{fkguo@itp.ac.cn}
\affiliation{CAS Key Laboratory of Theoretical Physics,  Institute of
Theoretical Physics, Chinese~Academy~of~Sciences,
 Zhong~Guan~Cun~East~Street~55,
Beijing~100190, China} \affiliation{School of Physical Sciences, University of Chinese Academy of
Sciences,\\ Beijing 100049, China}

\author{J.~Nieves}
\email{jmnieves@ific.uv.es}
\affiliation{Instituto de F\'{\i}sica Corpuscular (centro mixto CSIC-UV), Institutos de Investigaci\'on de Paterna,
Apartado 22085, 46071, Valencia, Spain}

\date{\today}
\begin{abstract}

By solving the Muskhelishvili--Omn\`es integral equations, the scalar form
factors of the semileptonic heavy meson decays $D\to\pi \bar \ell \nu_\ell$,
$D\to \bar{K} \bar \ell \nu_\ell$, $\bar{B}\to \pi  \ell \bar\nu_\ell$ and
$\bar{B}_s\to K  \ell \bar\nu_\ell$ are simultaneously studied.
As input, we employ unitarized heavy meson--Goldstone boson chiral coupled-channel amplitudes for
the energy regions not far from thresholds, while, at high energies,  adequate asymptotic conditions are imposed.  The scalar form factors are expressed  in
terms of  Omn\`es matrices multiplied by vector polynomials, which contain some undetermined dispersive subtraction constants. We make use of heavy quark and chiral symmetries to constrain these constants, which are fitted to  lattice QCD
results both in the charm and the bottom sectors, and in this latter sector to the light-cone sum rule predictions close to $q^2=0$ as well.  We find a good  simultaneous  description of the scalar form factors for the four  semileptonic decay reactions.  From this combined fit, and taking advantage  that scalar and vector form factors are equal at $q^2=0$, we obtain
$|V_{cd}|=0.244\pm 0.022$, $|V_{cs}|=0.945\pm 0.041$ and
$|V_{ub}|=(4.3\pm 0.7)\times10^{-3}$ for the involved Cabibbo--Kobayashi--Maskawa (CKM) matrix elements. In addition, we predict the following vector form factors at $q^2=0$:
$|f_+^{D\to\eta}(0)|=0.01\pm 0.05$, 
$|f_+^{D_s\to K}(0)|=0.50 \pm 0.08$, 
$|f_+^{D_s\to\eta}(0)|=0.73\pm 0.03$ and 
$|f_+^{\bar{B}\to\eta}(0)|=0.82 \pm 0.08$, 
which might serve as alternatives to determine the CKM elements when experimental measurements of the corresponding differential decay rates become available. Finally, we predict the different form factors above the $q^2-$regions accessible in the semileptonic decays, up to moderate energies amenable to be described using the 
unitarized coupled-channel chiral approach.
\end{abstract}

\maketitle
\tableofcontents

\section{Introduction}

Exclusive semileptonic decays play a prominent role in the precise determination of
the Cabibbo--Kobayashi--Maskawa~(CKM) matrix elements, which are particularly
important to test the standard model~(SM)---any violation of the unitarity
of the CKM matrix would reveal new physics beyond the SM (see for instance the review on the CKM mixing parameters by the Particle Data Group (PDG)~\cite{Olive:2016xmw}). Experimental and theoretical efforts have been devoted to multitude of
inclusive and exclusive semileptonic decays driven  by electroweak charge currents. For instance, the $K_{\ell3}$ decays and those of the type $H\to{\phi}\,\bar{\ell}\,\nu_\ell$ and $H\to{\phi}\,{\ell}\,\bar \nu_\ell$
(hereafter denoted by $H_{\ell3}$ or $H\to \phi$), where
$H\in\{D,\, \bar{B}\}$ is an open heavy-flavor pseudoscalar meson and $\phi \in\{\pi,\, K,\, \bar{K},\, \eta\}$ denotes one of the Goldstone bosons due to the spontaneous breaking of the approximate chiral symmetry of Quantum Chromodynamics (QCD),
are important in the extraction of some of the CKM  matrix elements. Experimentally,
significant progresses have been achieved and  absolute decay
branching fractions and differential decay rates have been accurately measured~\cite{Athar:2003yg, Hokuue:2006nr, Aubert:2006px, Aubert:2006ry, Widhalm:2006wz, Besson:2009uv,delAmoSanchez:2010af, Ha:2010rf, Lees:2012vv, Lees:2014ihu}. On the theoretical side,
determinations of the form factors in the vicinity of $q^2=0$ (with $q^2$ the invariant mass of the outgoing lepton pair) using light-cone sum rules (LCSR) have significantly improved their precision~\cite{Duplancic:2008ix,Duplancic:2008tk}, and have reached the level
of two-loop accuracy~\cite{Bharucha:2012wy}. Meanwhile, improvements have been made by using
better actions in lattice QCD (LQCD), which have allowed to extract CKM matrix elements with significantly reduced statistical and systematical uncertainties~\cite{Dalgic:2006dt, Na:2010uf,Na:2011mc,Bouchard:2014ypa,Lattice:2015tia,Flynn:2015mha,Lubicz:2017syv}. As a result of this  activity in the past decade,  lattice calculations on the scalar form
factors in heavy-to-light semileptonic transitions have been also reported by the different groups (see the informative review by the Flavour Lattice
Averaging Group (FLAG)~\cite{Aoki:2016frl}).

The extraction of the CKM mixing parameters from $K_{\ell3}$ and/or
$H_{\ell3}$ decays relies on the knowledge of the vector [$f_+(q^2)$] and scalar
[$f_0(q^2)$] hadronic form factors  that determine the matrix elements of the charged current between the initial and final hadron states\footnote{The contribution of the scalar form factor to the decay width is suppressed since it vanishes in the limit of massless leptons. However, both scalar and vector form factors take the same value at $q^2=0$, and thus an accurate determination of the $q^2-$dependence of the scalar form factor can be used to constrain the vector one in this region.}. Various parameterizations, such as the Isgur--Scora--Grinstein--Wise updated
model~\cite{Scora:1995ty} or  the series expansion proposed in Ref.~\cite{Becher:2005bg}, are
extensively used in LQCD  and  experimental studies. In this work, we will study the scalar form factors in
 $H_{\ell3}$ decays by using the Muskhelishvili--Omn{\`e}s (MO) formalism,
which is a model independent approach to account for $H \phi$
coupled-channel re-scattering effects. The coupled-channel MO formalism has been extensively applied to the scalar
$\pi\pi$, $\pi K$ and $\pi\eta$ form factors, see, e.g., Refs~\cite{Donoghue:1990xh,Jamin:2001zq,Albaladejo:2015aca,Daub:2015xja, Albaladejo:2016mad}. It builds up an
elegant bridge to connect the form factors with the corresponding $S$-wave
scattering amplitudes via dispersion relations. The
construction of those equations is rigorous in the sense that the fundamental
principles, such as unitarity and analyticity, and the proper QCD asymptotic
behaviour are implemented. The first attempts to extend this method to the
investigation of the scalar $H\to  \phi$ form factors were made in
Refs.~\cite{Burdman:1996kr, Flynn:2000gd, Albertus:2005ud, Flynn:2006vr, Flynn:2007ki, Flynn:2007qd, Flynn:2007ii, Meissner:2013pba, Albertus:2014gba}, but just for the single-channel case. A similar dispersive MO approach has been also employed
to study the semileptonic  $\bar B \to \rho l \bar \nu_l$ ~\cite{Flynn:2008zr,Meissner:2013pba, Kang:2013jaa, Albertus:2014xwa}  and $\bar B_s \to \bar K^* l \bar \nu_l$ \cite{Meissner:2013pba} decays  and the possible extraction of the CKM element $|V_{ub}|$ from data on the four-body $\bar B \to \pi\pi l \bar \nu_l$  and $\bar B_s \to \bar K \pi l \bar \nu_l$ decay-modes. 

The study of heavy-light form factors using the MO representation incorporating coupled-channel effects has not been undertaken yet. 
This is mainly because of the poor knowledge on the $H\phi$ interactions up to very recent years. However, a few intriguing positive-parity charmed mesons, like the $D_{s0}^\ast(2317)$, have been recently discovered~\cite{Olive:2016xmw}, giving support to a new paradigm for heavy-light meson spectroscopy~\cite{Du:2017zvv} that questions their traditional $q\bar q$  constituent quark model interpretation. Hence, the study  of the $H\phi$
interactions aiming at understanding the dynamics of these newly observed states has become an 
interesting subject by itself, see, e.g.,  Refs.~\cite{Kolomeitsev:2003ac,Hofmann:2003je,Guo:2006fu,Gamermann:2006nm,Guo:2009ct,Wang:2012bu,Altenbuchinger:2013vwa,Yao:2015qia,Guo:2015dha,Albaladejo:2016lbb,Du:2017ttu,Guo:2018kno} 
for phenomenological studies  and \cite{Liu:2008rza,Liu:2012zya,Mohler:2012na,Mohler:2013rwa,Lang:2014yfa,Moir:2016srx,Bali:2017pdv} for LQCD calculations. For the $D_{\ell3}$ decays, several  LQCD results on the relevant form factors have been recently reported, see, e.g., Refs.~\cite{Na:2010uf,Na:2011mc,Lubicz:2017syv}. This situation makes timely the study of the scalar $D\to  \phi$ form
factors by means of the MO representation  incorporating  our current knowledge
of $D\phi$ interactions. The extension to the $H=\bar{B}$ case is straightforward with
the help of heavy quark flavour symmetry (HQFS). Based on HQFS, the low energy constants (LECs) 
involved in the $D\phi$ interactions or $D\to \phi$ semileptonic
form factors are related to their analogues in the bottom sector by specific
scaling rules.
It is then feasible either to predict quantities in the bottom (charm) sector by making
use of the known information in the charm (bottom) case or to check how well HQFS works
by testing the scaling rules.

In the present study, we construct the MO representations of the scalar form
factors, denoted by $f_0(s)$, for the semileptonic $D\to\pi$ and $D\to \bar{K}$
 transitions, which are related to the unitarized $S$-wave
scattering amplitudes in the $D\phi$ channels with strangeness~($S$) and isospin~($I$) quantum numbers
$(S,I)=(0,\frac{1}{2})$ and $(S,I)=(1,0)$, respectively. These amplitudes are obtained by unitarizing the $\mathcal{O}(p^2)$ heavy-meson chiral perturbative ones~\cite{Guo:2008gp}, with LECs  determined from  the
lattice calculation ~\cite{Liu:2012zya} of the $S$-wave scattering lengths in several $(S,I)$ sectors. The scheme provides an accurate description of the $D\phi$ interactions in coupled channels. For instance, as it is shown in Ref.~\cite{Albaladejo:2016lbb}, the finite volume energy levels in the $(S,I)=(0,1/2)$ channel calculated with the unitarized amplitudes, without adjusting any parameter, 
are in an excellent agreement with those recently reported by  the Hadron Spectrum Collaboration~\cite{Moir:2016srx}. In addition,  it is demonstrated in Ref.~\cite{Du:2017zvv} that these well constrained amplitudes for Goldstone bosons scattering off charm mesons are fully consistent with recent high quality data on the $B^- \to D^+\pi^-\pi^-$ final states provided by the LHCb experiment~\cite{Aaij:2016fma}.

The unitarized chiral
scattering amplitudes are used in this work as inputs to the dispersive integrals. However,
these amplitudes are valid only in the low Goldstone-boson energy region. Hence, asymptotic behaviors at high energies for the phase shifts and inelasticities
are imposed in the solution of the MO integral equations. The Omn\`es matrices obtained in this way incorporate the strong final state interactions, and the scalar form factors are calculated by multiplying the former by polynomials. The ({\it a priori} unknown) coefficients of the polynomials are expressed in terms of the LECs appearing at next-to-leading order (NLO) in the chiral expansion of the form factors~\cite{Wise:1992hn,Burdman:1992gh}. 

The scheme employed in the charm sector is readily extended to the bottom one. Afterwards, the LECs could be either 
determined by fitting to the results obtained  in the LQCD analyses of the
$D\to\pi(\bar{K})$ decays carried out in Refs.~\cite{Na:2010uf,Na:2011mc, Lubicz:2017syv} or  to the LQCD and LCSR combined $\bar{B}\to\pi$ and  $\bar{B}_s\to K$  scalar form factors reported in Refs.~\cite{Dalgic:2006dt,Bouchard:2014ypa,Lattice:2015tia,Flynn:2015mha} and \cite{Duplancic:2008ix,Duplancic:2008tk,Bharucha:2012wy}, respectively. 
In both scenarios LQCD and LCSR results are well described using  the MO dispersive representations of the scalar form factors constructed in this work.

However, our best results are obtained by a simultaneous fit to all available results, both in the charm and bottom sectors.  As mentioned above, all of the
LECs involved in the $\bar{B}_{(s)}\phi$ interactions or $\bar{B}_{(s)}\to\bar{\phi}$ semileptonic transitions are related to those in the charm sector by making use
of the heavy quark scaling rules~\cite{Burdman:1992gh}, which introduce some constraints between the polynomials that appear in the different channels. Thus, assuming a reasonable effect of the HQFS
breaking terms, a combined fit is performed to the $D\to
\pi/\bar{K}$ and the  $\bar{B}\to\pi$ and  $\bar{B}_s\to K$ scalar form factors, finding  a fair description of all the fitted data, and providing reliable predictions of  the different scalar form factors in
the whole semileptonic decay phase space, which turn out to be compatible with other theoretical determinations by, e.g., perturbative QCD~\cite{Li:2012nk,Wang:2012ab}. The results of the fit allow also to predict  
the scalar form factors for the $D\to\eta$, $D_s\to {K}$ and $D_s\to \eta$ transitions in the charm sector,  and for the very first time for the $\bar{B}\to\eta$ decay. In some of these transitions, the form factors  are difficult for LQCD due to the existence of disconnected diagrams of quark loops.

Based on the results of the combined fit, and  taking advantage of the fact that scalar and vector form factors are equal at $q^2=0$, we extract all the heavy-light CKM elements and test the second-row unitarity by using the $|V_{cb}|$ value given in the PDG~\cite{Olive:2016xmw}. We also predict the different form factors above the $q^2-$regions accessible in the semileptonic decays, up to energies in the vicinity of the involved thresholds, which 
should be correctly described within the employed  unitarized chiral approach.

This work is organized as follows. In Subsec.~\ref{sec:FF}, we introduce the
definitions of the form factors for $H_{\ell3}$ decays. A general overview of the
MO representation of the scalar form factors is given in Subsec.~\ref{sec:MO}, while 
the inputs for the MO problem and the solutions are discussed in
Subsec.~\ref{sec:MOsol}. In Subsec.~\ref{sec:SC}, we derive the scalar as well
as vector form factors at NLO in heavy-meson chiral perturbation theory and then
perform the aforementioned matching between the MO and the chiral representations close to the corresponding thresholds. Section~\ref{sec:NR} comprises our numerical results and discussions, with details of the fits given in  Subsecs.~\ref{sec:fit} and \ref{sec:combinedfit}. With the
results of the combined charm-bottom fit, in
Sec.~\ref{sec:CKM}, we extract the related CKM elements and make
predictions  for the values of $f_+(0)$ in various transitions. Predictions  
for flavour-changing $b\to u$ and $c\to d, s$ scalar form factors above the $q^2-$regions accessible in the  semileptonic decays 
are given and discussed in Subsec.~\ref{sec:sfq2max}.
We summarize the results of this work in Sec. ~\ref{sec:CON}. Finally,  
the heavy-quark scaling rules of the LECs involved in the $H\phi$ interactions are 
discussed in  Appendix~\ref{sec:scaling.h}, while some further results for $b\to u$ form factors, obtained with quadratic MO polynomials, are shown in Appendix~\ref{sec:quadMo}.

\section{Theoretical framework\label{sec:theo}}

\subsection[Form factors in $H_{\ell3}$ decays]{\boldmath Form factors in $H_{\ell3}$ decays\label{sec:FF}}

For a semileptonic decay of the type
$H(p)\to\bar{\phi}(p^\prime)\,{\ell}(p_\ell)\,\bar{\nu}_\ell(p_\nu)$, the 
Lorentz invariant Feynman amplitude is proportional to 
\begin{align}
\mathcal{M} \propto &
\frac{G_F}{\sqrt{2}}\Big\{\bar{u}(p_\ell)\gamma^\mu(1-\gamma_5)v(p_\nu)\Big\} \times
\nonumber\\
& \Big\{V_{Qq}\,\langle\bar{\phi}(p^\prime)|\bar{q}\gamma_\mu(1-\gamma_5)Q|H(p)
\rangle\Big\}\ , 
\end{align}
where $G_F$ is the Fermi constant, and $V_{Qq}$ is the CKM
matrix element corresponding to the flavour changing $Q\to q$ transition. The terms in the
first and second curly brackets stand for the weak and  hadronic matrix elements,
respectively. In the hadronic matrix element, the axial-vector part vanishes
due to the parity conservation, while the remaining vector part is
parametrized in a conventional form as
\begin{align}
\langle\bar{\phi}(p^\prime)|\bar{q}\gamma^\mu{Q}|H(p)\rangle= & 
f_+(q^2)\bigg[\Sigma^{\mu}-\frac{m_H^2-M_\phi^2}{q^2}q^\mu\bigg] \nonumber\\
& + f_0(q^2)\frac{m_H^2-M_\phi^2}{q^2}q^\mu~, 
\end{align}
where $f_+(q^2)$ and $f_0(q^2)$ are the vector and scalar form factors,
respectively, with $q^\mu=p^\mu-p^{\prime\mu}$ and
$\Sigma^\mu=p^\mu+p^{\prime\mu}$. Note that both form factors should be equal at $q^2=0$. 
As discussed in Ref.~\cite{Gasser:1984ux}, they specify the $P$-wave ($J^P=1^-$)
and $S$-wave ($J^P=0^+$) of the crossed-channel matrix elements,
\bea
\label{eq:ffcrossing}
\langle0|\bar{q}\gamma^\mu{Q}|H(p){\phi}(-p^\prime)\rangle=\langle\bar{\phi}(p^\prime)|\bar{q}\gamma^\mu{Q}|H(p)\rangle\
.
\eea
Both the scalar and vector form factors contribute to the differential decay
rate, see e.g., Ref.~\cite{Aoki:2016frl}. Nevertheless, when the lepton mass is
neglected, the differential decay rate in the $H-$meson rest frame can be simply expressed in terms of the
vector form factor via 
\bea
\label{eq:decay} 
\frac{{\rm
d}\Gamma(H\to\bar{\phi}\ell\bar{\nu}_\ell)}{{\rm
d}q^2}=\frac{G_F^2}{24\pi^3}|\vec{p}^{\,\prime}|^3|V_{Qq}|^2|f_+(q^2)|^2~. 
\eea
It is then possible to extract the CKM element $|V_{Qq}|$ even for a single value of the four-momentum transfer, provided one simultaneously knows the vector form factor and the experimental differential decay width. A possible choice is $q^2=0$, where the scalar and vector  form factors
coincide, $f_+(0)=f_0(0)$. 

In the next two subsections, Subsecs.~\ref{sec:MO}--\ref{sec:MOsol}, we give specific details on the MO representation of the form factors. 
For brevity, in some occasions  we will focus on the formalism for the case of the charm
sector ($H=D$). The extension to the bottom sector ($H=\bar{B}$) is straightforward using HQFS, though some aspects are explicitly discussed in Subsec.~\ref{sec:ext-bottom}.

\subsection{Muskhelishvili--Omn{\`e}s representation\label{sec:MO}}

We now discuss the dispersive representation of the
scalar form factors within  the MO formalism.  Throughout this work, isospin breaking terms are not considered, and therefore it
is convenient to work with the isospin basis. Before proceeding, we first discuss the relation of the form factors expressed in the particle and isospin bases. We start defining the phase convention for isospin states:
\bea
|D^+\rangle &=& -\left|\frac{1}{2},+\frac{1}{2}\right\rangle\ ,\qquad
|\pi^+\rangle = -|1,+1\rangle\ ,\nonumber\\
 |\bar{K}^0\rangle &=& -\left|\frac{1}{2},+\frac{1}{2}\right\rangle\ , \qquad
 |\bar{B}^0\rangle = -\left|\frac{1}{2},+\frac{1}{2}\right\rangle\ ,
\eea
while the other states are defined with a positive sign in front of the $|I,I_3\rangle$ state. The form factors involving the $c \to d$ transition are those appearing in the $D^0 \to \pi^-$, $D^+ \to \pi^0$, $D^+\to \eta$, and $D_s^+ \to K^0$ semileptonic decays. (Note that $D^0 \to \pi^-$ and $D^+ \to \pi^0$ transitions are related by an isospin rotation.) The details of these form factors close to the zero recoil point, where the outgoing Goldstone boson is at rest, are greatly influenced by the $\pi D$, $D\eta $, and $D_s \bar{K}$  scattering amplitudes in the $(S,I)=(0,1/2)$ sector. Note that to be consistent with the convention of Refs.~\cite{Guo:2008gp,Liu:2012zya}, we use $\pi D$ instead of $D\pi$  to construct the isospin 1/2 state.
We then define the vector-column  $\vec{\cal{F}}^{(0,1/2)}$ as
\begin{equation}\label{eq:mo.dpi} 
\vec{\cal{F}}^{(0,1/2)}(s) \equiv \left(
\begin{array}{c}
  \sqrt{\frac{3}{2}}  f_0^{D^0\to \pi^-}(s) \\
f_0^{D^+\to{\eta}}(s)  \\
f_0^{D_s^+\to{K^0}}(s)
\end{array}
\right)~.
\end{equation}
We shall use the shorthands $f_0^{D\pi}(s) = f_0^{D^0 \to \pi^-}(s)$, $f_0^{D\eta}(s) = f_0^{D^+ \to \eta}(s)$, and $f_0^{D_s \bar{K}}(s) = f_0^{D^+_s \to K^0}(s)$. Likewise for the $c \to s$ transitions, we have the  $D^0 \to K^-$, $D^+ \to \bar{K}^0$, and $D^+_s \to \eta$ semileptonic decays, related to the $DK$ and $D_s \eta$ scattering amplitudes in the $(S,I)=(1,0)$ sector. We thus define\footnote{Here also  $D^0 \to K^-$ and $D^+ \to \bar{K}^0$ form factors are related by an isospin rotation. }:
\begin{equation}\label{eq:mo.dk}
\vec{\cal{F}}^{(1,0)}(s) \equiv \left(
\begin{array}{c}
  -\sqrt{2}  f_0^{D^0\to K^-}(s) \\
f_0^{D_s^+\to{\eta}}(s) 
\end{array}
\right)~,
\end{equation}
for which we will also use the notation $f_0^{D^0 \to K^-}(s) = f_0^{DK}(s)$ and $f_0^{D^+_s \to \eta}(s) = f_0^{D_s \eta}(s)$. Here, and again to be consistent with the convention of Refs.~\cite{Guo:2008gp,Liu:2012zya}, we use $D K$ instead of $K D$  to construct the isoscalar state, which is just the opposite convention to that used for the isospin $\pi D$ state. 
With these definitions, the unitarity relation for any of the $\mathcal{\vec{F}}(s)$ can be compactly written as:
%
\begin{eqnarray}
\frac{\vec{\cal F}(s+i\epsilon)-\vec{\cal F}(s-i\epsilon)}{2i}&=&{\rm Im}\,\vec{\cal F}(s+i\epsilon)\nonumber  \\
&=& T^\ast(s+i\epsilon)\Sigma(s)\vec{\cal F}(s+i\epsilon)\ , \label{eq:unita}
\end{eqnarray}
%
where $T(s)$ stands for the coupled-channel $S$-wave scattering amplitude in the corresponding $(S,I)$ sector, which will be discussed further in Sec.~\ref{sec:MOsol}. The diagonal matrix  $\Sigma(s)$  contains the phase space factors. For $(0,1/2)$, one has $\Sigma(s) = \text{diag}\left(\sigma_{D\pi}(s), \sigma_{D\eta}(s), \sigma_{D_s \bar{K}}(s) \right)$, whereas in the $(1,0)$ sector $\Sigma(s) = \text{diag} \left( \sigma_{DK}(s), \sigma_{D_s\eta} \right)$. The function $\sigma_{ab}(s)$ is defined as:
\begin{equation} 
\sigma_{ab}(s)=\frac{\lambda^{1/2}(s,m_a^2,m_b^2)}{s}\Theta\left[s-(m_a+m_b)^2\right]~,
\end{equation}
with $\lambda(x,y,z)=x^2+y^2+z^2-2xy-2yz-2xz$, the K{\"a}hl\'en function. Invariance under time-reversal together with  the optical-theorem leads to 
\begin{equation}
 {\rm Im}\, T^{-1} (s + i \epsilon) = - \Sigma (s) \label{eq:defImTinv}
\end{equation}
In this convention, the $T$- and $S$-matrices are related by
\begin{align}
S(s) = \mathbb{I}+2i \Sigma^\frac12 (s) T(s) \Sigma^\frac12 (s)  \label{eq:defS} 
\end{align}
The unitarity relation of Eq.~\eqref{eq:unita} can be used to obtain dispersive representations for the form factors. We start considering the $D\to\pi$ transition in the single channel case (elastic unitarity) where the form factor satisfies
\begin{equation}
\label{eq:single} 
{\rm Im} f_0^{D\pi}(s) = [t^{\pi D}(s)]^\ast \sigma_{D\pi}(s) f_0^{D\pi}(s)\ ,\quad s\equiv q^2~,
\end{equation}
with $t^{\pi D}(s)$ the $\pi D$ $S$-wave elastic scattering amplitude. Equation~\eqref{eq:single} admits an algebraic solution~\cite{Omnes:1958hv},
\begin{equation}
f_0^{D\to\pi}(s)= \Omega(s) P(s)\ ,
\end{equation}
where $P(s)$ is an undetermined polynomial, and the Omn{\`e}s function $\Omega(s)$ is given by
\begin{equation}
\label{eq:OmnesElastic}
\Omega(s)=\exp\bigg[\frac{s}{\pi}\int_{(m_D+M_\pi)^2}^{\infty}{\rm d}s^\prime\frac{\delta(s^\prime)}{s^\prime(s^\prime-s)}\bigg]\ ,
\end{equation}
with $\delta(s)$ the elastic $0^+$ $\pi D$ phase shift, in accordance with the
Watson final state interaction theorem~\cite{Watson:1954uc}. This was the scheme adopted in the previous studies carried out in Refs.~\cite{Burdman:1996kr, Flynn:2000gd, Albertus:2005ud, Flynn:2006vr, Flynn:2007ki, Flynn:2007qd, Flynn:2007ii, Albertus:2014gba}.

For coupled channels the solution for $\mathcal{\vec{F}}(s)$ takes the form:
\bea
\label{eq:FFmatrix}
\vec{\cal F}(s)=\Omega(s)\cdot\vec{\cal P}(s) \,
\eea
being $\vec{\cal P}(s)$ a vector of polynomials with real coefficients and $\Omega(s)$ the Omn{\`e}s matrix that satisfies\footnote{Taking into account that the $\Omega(s)$ matrix should have only a right-hand cut and it should be real below all thresholds, Eq.~\eqref{eq:Omega} is equivalent to 
\begin{equation}
 \Omega(s+i\epsilon) = H(s+i\epsilon)\Omega(s-i\epsilon).\label{eq:eq1}
\end{equation}
with $H(s)= \left(\mathbb{I}+ 2i T(s)\Sigma(s)\right)$. Furthermore since $H(s)H^\ast(s)=\mathbb{I}$, though $H(s)$ is not the $S$-matrix in the coupled-channel case, 
it follows
\begin{equation}
 {\rm det}\left[\Omega(s+i\epsilon)\right] = e^{2i\phi(s)}{\rm det}\left[\Omega(s-i\epsilon)\right] \label{eq:eq2}
\end{equation}
with $\exp{2i\phi(s)}={\rm det}\left[H(s)\right]$. This is to say that the determinant of the matrix $\Omega(s)$ satisfies a single-channel Omn{\`e}s-type relation~\cite{Moussallam:1999aq}, which is extensively used in this work to check  the accuracy of the numerical calculations. Note that above all thresholds, ${\rm det}\left[H(s)\right]= {\rm det}\left[S(s)\right]$ and therefore in the elastic case ($\eta_i \to 1$ $\forall i$), $\phi(s)=\sum_{i=1}^n, \delta_i(s)$,
with $n$ the number of channels.}
\begin{equation}
 {\rm Im}\, \Omega(s+i\epsilon) = T^*(s+i\epsilon)\Sigma(s)\Omega(s+ i\epsilon) \label{eq:Omega}
\end{equation}
which leads to the following unsubtracted dispersion relation:
\bea
\label{eq:OmnesUDR}
\Omega(s+i\epsilon)=\frac{1}{\pi}\int_{s_{\rm th}}^\infty\frac{T^\ast(s^\prime)\,\Sigma(s^\prime)\,\Omega(s^\prime)}{s^\prime-s-i\epsilon}{\rm d}s^\prime\ ,
\eea
with $s_{\rm th}$ the lowest threshold,
which is referred to as the MO integral equation~\cite{Muskhelishvili}. Taking a polynomial $\vec{P}(s)$ of rank $(n-1)$ would require the knowledge of $\mathcal{\vec{F}}(s)$ for $n$ values of $s$. Unlike the single channel case, there is no analytical solution [in the sense of Eq.~\eqref{eq:OmnesElastic}] for the coupled-channel MO problem and it has to be solved numerically. The MO equation can be written in an alternative form: 
\bea
\label{eq:MOpro} 
{\rm Re}\,\Omega(s)&=&\frac{1}{\pi}\mathcal{P}\int_{s_{\rm th}}^\infty
\frac{{\rm d}s^\prime}{s^\prime-s} {\rm Im}\,\Omega(s'), \nonumber\\
{\rm Im}\,\Omega(s)&=&X(s)\,{\rm Re}\,\Omega(s)\ ,
\eea
with $\mathcal{P}\int$ denoting the principal value and 
\bea
X(s)={\rm Im}\,T(s)\left[{\rm Re}\,T(s)\right]^{-1}\ ,
\eea
which is expressed in terms of the $T$-matrix and encodes the
information on the $D\phi$ re-scattering.  The linear MO system,
Eq.~(\ref{eq:MOpro}), can be solved by following the appropriate numerical 
method described in Ref.~\cite{Moussallam:1999aq}.

\subsection{Inputs and MO solutions \label{sec:MOsol}}

To solve the MO equation and obtain the $\Omega(s)$ matrix, the $T$-matrix is needed as an input. We will use here the amplitudes based on
unitarized chiral effective theory as computed in Refs.~\cite{Guo:2008gp,Liu:2012zya}. Because of the normalizations used in Eqs.~\eqref{eq:unita} and \eqref{eq:defImTinv}, the unitarized amplitude in Ref.~\cite{Liu:2012zya}, denoted here by $T_U(s)$, is related to the $T$-matrix introduced in the above subsection by:
\begin{equation}
T(s)=-\frac{1}{16\pi}T_U(s)~.
\end{equation}
This unitary $T$-matrix is written as:
\begin{equation}
T_U^{-1}(s) = V^{-1}(s) - G(s)~,  \label{eq:deftu}
\end{equation}
where the elements of the diagonal matrix $G(s)$ are the two-meson loop functions~\cite{Liu:2012zya}, and the matrix $V(s)$ contains the interaction {\it potentials} which, as pointed out above, are taken from the $\mathcal{O}(p^2)$ chiral Lagrangians deduced in Ref.~\cite{Guo:2008gp} (see Ref.~\cite{Albaladejo:2016lbb} for more details).\footnote{The most recent unitarized amplitudes based on one-loop potentials are given in
Refs.~\cite{Yao:2015qia,Du:2017ttu}. However, we will not use them here since the LECs involved in the one-loop analyses can not be well determined 
due to the lack of precise data, as pointed out in
Ref.~\cite{Du:2016tgp}.} In Ref.~\cite{Liu:2012zya} all the parameters involved in the $T$-matrix, which are a few LECs and one subtraction constant, have been pinned down by fitting to the
lattice calculation of the $S$-wave scattering lengths in several $(S,I)$ sectors. The obtained interaction potentials successfully describe~\cite{Albaladejo:2016lbb} the  charm $(S,I)=(0,1/2)$  finite volume energy levels, without adjusting any parameter, calculated by the Hadron Spectrum Collaboration~\cite{Moir:2016srx}. Moreover, as mentioned in the Introduction, these well constrained amplitudes turned out to be fully consistent~\cite{Du:2017zvv} with the LHCb data on the $B^- \to D^+\pi^-\pi^-$ reaction~\cite{Aaij:2016fma}.

Since the amplitudes are based on chiral potentials, the obtained $T$-matrices are only valid in the energy region not far from the corresponding thresholds. We thus adopt such $T$-matrices only up to a certain value
of $s$, denoted by ${s}_m$. Above that energy, the $T$-matrix elements are computed as an interpolation between their values at $s=s_m$ and the asymptotic values at $s=\infty$. The interpolation still gives a unitary $T$-matrix since, as we will specify below, it is actually performed on the phase shifts and the inelasticities. Moreover, the approximation of quasi-two-body channels 
cannot hold for arbitrarily large energies and Eq.~\eqref{eq:unita} is a
reasonable approximation to the exact discontinuity only
in a finite energy range. However, as we are interested in constructing the form factors in a finite energy region also, the detailed behaviour of the spectral function
at much higher energies should be, in principle, unimportant. As we will see below, this is not entirely correct, in particular for the $\bar B\to \pi$ semileptonic transition, because of the large 
$q^2-$phase space accessible in this decay. Nevertheless, we will assume that  Eq.~\eqref{eq:unita}  holds up to infinite energies, only
requiring that the $T$-matrix behaves in a way that ensures
an appropriate asymptotic behaviour of the form factors, and we will discuss the dependence of our results on the contributions from the high energy region. In general, except for the $\bar B$ decays\footnote{We will specifically discuss the situation for these transitions below.}, the asymptotic conditions on the $T$-matrix are chosen such that:
\begin{align}
\lim_{s\to \infty} \lvert T_{ij}(s) \rvert & = 0\quad \text{for}\quad i \neq j~,\label{eq:deltas}\\
\lim_{s\to \infty} \sum_{i=1}^n \delta_i(s) & = n\pi~,\label{eq:npi}
\end{align}
where $n$ is the number of channels involved in the $T$-matrix and $\delta_i(s)$ are the phase shifts. These conditions ensure (in general) that the unsubtracted dispersion relation for the Omn\`es matrix in Eq.~\eqref{eq:OmnesUDR} has a unique solution, albeit a global normalization \cite{Muskhelishvili} (see also in particular Sec.~4.3 of Ref.~\cite{Moussallam:1999aq}). 
The condition of Eq.~\eqref{eq:npi} guaranties that 
\begin{equation}
 \lim_{s\to \infty} {\rm det}\left[\Omega(s)\right] \to 1/s^n \label{eq:assydet}
\end{equation}
as can be deduced from the discussion of Eqs.~\eqref{eq:eq1} and ~\eqref{eq:eq2}. Note that the normalization of the Omn\`es matrix is completely arbitrary, and the computed form factors do not depend on it\footnote{For example, let us consider Omn\`es matrices $\Omega$ and $\bar{\Omega}$  normalized to $\Omega(0) = \mathbb{I}$ or $\bar{\Omega}(s_n) = A$ ($s_n$ the normalization point, $s_n \leqslant s_\text{th}$ and $A$ a real matrix), respectively. The matrix $\bar{\Omega}(s)$ is readily obtained from $\Omega(s)$ as $\bar{\Omega}(s) = \Omega(s) \Omega^{-1}(s_n) A$. The form factors can then also be written as $\mathcal{\vec{F}}(s) = \Omega(s) \mathcal{\vec{P}}(s) = \bar{\Omega}(s) A^{-1} \Omega(s_n) \mathcal{\vec{P}}(s) \equiv \bar{\Omega}(s) \mathcal{\vec{\bar{P}}}(s)$, where the definition of $\mathcal{\vec{\bar{P}}}(s)$ reabsorbs the constant matrix $A^{-1} \Omega(s_n)$.}.

In what follows,  we detail the $T$-matrices and the specific shape of the asymptotic conditions for the two coupled-channel cases [$(S,I)=(1,0)$ and $(0,1/2)$] that will be analyzed below.

\subsubsection{The $(S,I)=(1,0)$ sector}
\begin{figure*}[t]
\begin{center}
\includegraphics[width=0.89\textwidth]{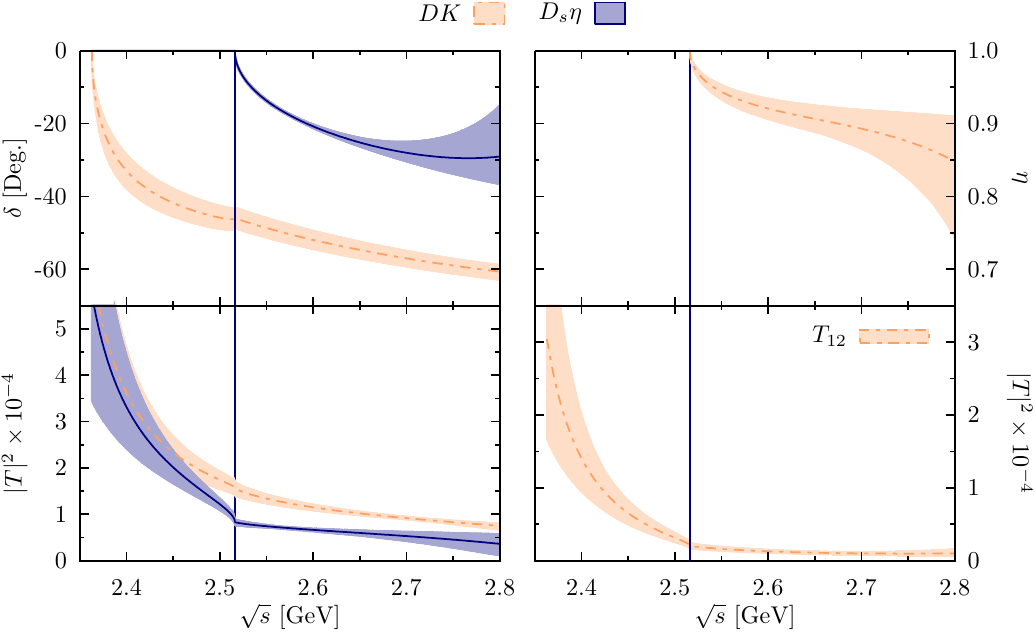}
\caption{Phase shifts, inelasticities [see Eq.~\eqref{eq:2defpheta}] and amplitude moduli from $T_{U}^{(1,0)}$. The vertical line indicates the $D_s \eta$ threshold. Error bands have been obtained by  Monte Carlo  propagating  the uncertainties of the LECs quoted in Ref.~\cite{Liu:2012zya}.\label{fig:in2cc}}
\end{center}
\end{figure*}
\begin{figure*}[t]
\begin{center}
\includegraphics[width=0.89\textwidth]{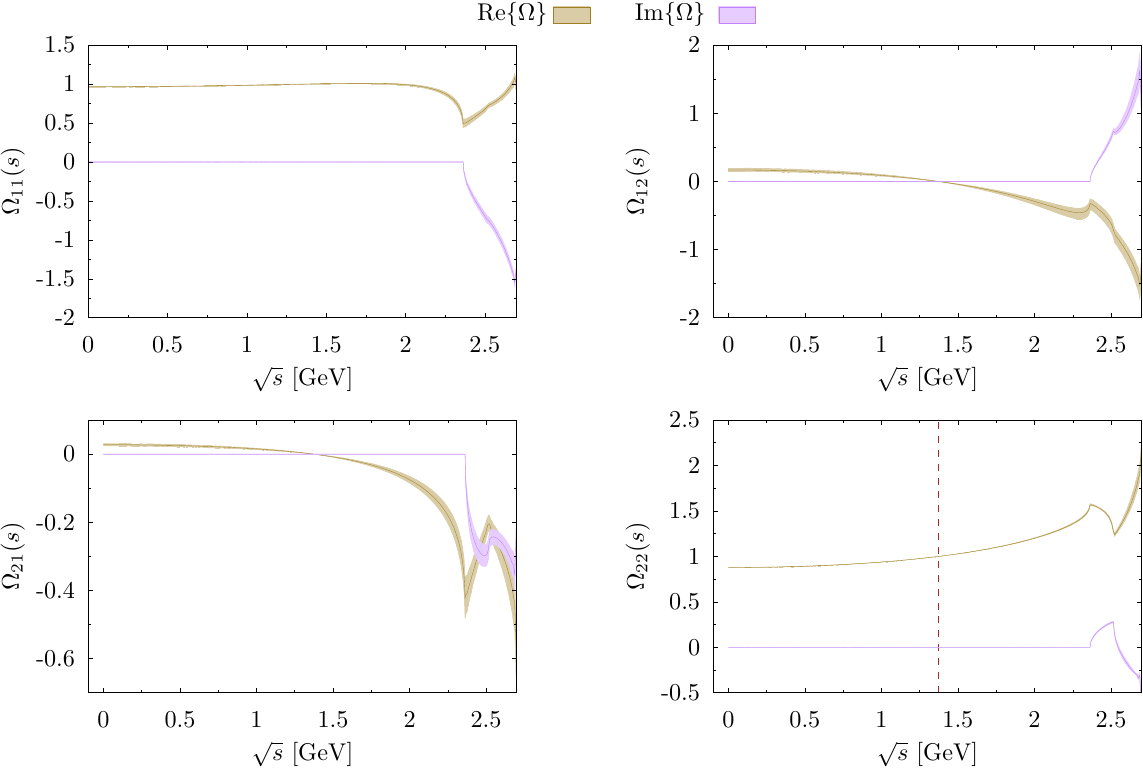}
\caption{$(S,I)=(1,0)$  Omn\`es matrix solution of the MO integral equation \eqref{eq:MOpro} with the contour condition $\Omega\left((m_D-M_K)^2\right) = \mathbb{I}$, and asymptotic  phase shifts $\delta_{DK}(\infty)=2\pi$ and $\delta_{D_s\eta}(\infty)=0$. Error bands have been obtained by  Monte Carlo  propagating  the uncertainties of the LECs quoted in Ref.~\cite{Liu:2012zya}. The dashed vertical line indicates $q^2_{\rm max}= (m_D-M_K)^2$. \label{fig:out2cc}}
\end{center}
\end{figure*}
In this sector, we will consider two coupled channels, $DK$ ($1$) and $D_s \eta$ ($2$), and above all thresholds, the $T$-matrix is parametrized in terms of two phase shifts and one inelasticity parameter, 
\bea
T(s)=
\left(
\begin{array}{cc}
\frac{\eta(s)e^{2i\delta_1}-1}{2i\sigma_1(s)}& \frac{\sqrt{1-\eta^2}e^{i\phi_{12}}}{2\sqrt{\sigma_1(s)\sigma_2(s)}}\\
\frac{\sqrt{1-\eta^2}e^{i\phi_{12}}}{2\sqrt{\sigma_1(s)\sigma_2(s)}}&\frac{\eta(s)e^{2i\delta_2}-1}{2i\sigma_2(s)} \label{eq:2defpheta}
\end{array}
\right)
\eea
with the phase $\phi_{12}=\delta_1+\delta_2+{\rm mod}(\pi)$ and $0\le \eta \le 1$.
To solve the MO integral equation [cf.~Eq.~\eqref{eq:MOpro}], we use a $T$-matrix of the form
\bea\label{eq:in2cc}
T(s)=\left\{
\begin{array}{cc}
 -\frac{1}{16\pi}T_{U}^{(1,0)}(s) &  s_{\rm th} \leqslant s \leqslant s_m\ ,     \\
 T_H(s) & s \geqslant s_m\ ,      \\
\end{array}
\right.
\eea
with $s_{\rm th}=(m_D+M_K)^2$ the lowest threshold, $T_U$ defined in Eq.~\eqref{eq:deftu} and $T_H$ the asymptotic matrix that will be discussed below.  Phase shifts, inelasticities and amplitude moduli from  $T_{U}^{(1,0)}$ are displayed in Fig.~\ref{fig:in2cc} up to $\sqrt{s}=2.8$ GeV, slightly above
$s_m=(2.7~{\rm GeV})^2$. Above this scale, the $T$-matrix elements are computed 
as an interpolation between their values at $s=s_m$ and the asymptotic values at $s=\infty$, given in Eqs.~\eqref{eq:deltas} and \eqref{eq:npi}. Thus,  $T_H$ is constructed from Eq.~\eqref{eq:2defpheta} using  
the following parameterizations for phase shifts and inelasticities:
\bea\label{eq:in2asy}
\delta_i(s)&=&\delta_i(\infty)+[\delta_i(s_m)-\delta_i(\infty)]\frac{2}{1+
({s}/{s_m})^{3/2}},\nonumber\\
\eta(s)&=&\eta(\infty)+[\eta(s_m)-\eta(\infty)]\frac{2}{1+({s}/{s_m})^{3/2}}\
\eea
as suggested in Ref.~\cite{Moussallam:1999aq}. As discussed above, the Omn\`es matrix is uniquely determined by choosing $\eta(\infty) = 1$ and $\delta_1(\infty) + \delta_2(\infty) = 2\pi$. The only remaining freedom is the distribution of $2\pi$ over the two phase shifts. Note that, $\delta_i(s_m)$ is defined modulo $\pi$ and this ambiguity  is fixed by continuity-criteria. 
Here for the $DK-D_s\eta$ coupled channels, we choose $\delta_{1}(\infty)=2\pi$, $\delta_2(\infty)=0$. Different 
choices of the asymptotic values or of the interpolating functions in Eq.~(\ref{eq:in2asy})
will modify the shape of the Omn\`es solution far from the chiral region. The numerical effect of such freedom on the derived 
scalar form factors should be safely compensated  by the undetermined polynomial in front of the Omn\`es matrix.

In Fig.~\ref{fig:out2cc}, we show the  solution of the 
MO integral equation \eqref{eq:MOpro}, with the input specified above,  and the contour condition $\Omega(q^2_{\rm max})=\mathbb{I}$, with $q^2_{\rm max}=(m_D-M_K)^2$.  We display results only up to $s=s_m$ that would be later used to evaluate the scalar form factors entering in the $D\to \bar K$  and $D_s \to \eta$ semileptonic transitions. Note that the imaginary parts are zero below the lowest threshold $s_{\rm th}=(m_D+M_K)^2$, and how the opening of the $D_s\eta $ threshold produces clearly visible effects in the  Omn\`es matrix. At very high energies, not shown in the figure, both real and imaginary parts of all matrix elements go to zero, as expected from Eq.~\eqref{eq:assydet}.

\subsubsection{The $(S,I)=(0,1/2)$ sector \label{sec:d012}}
\begin{figure*}[t]
\begin{center}
\includegraphics[width=0.89\textwidth]{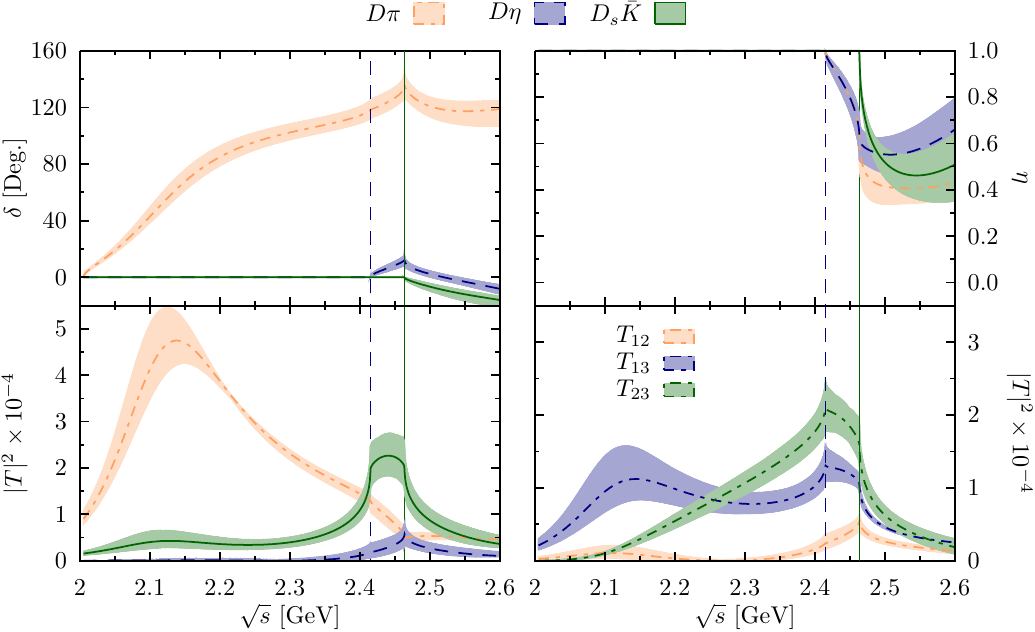}
\caption{Phase shifts, inelasticities  and amplitude moduli from $T_{U}^{(0,1/2)}$ in the charm sector. The vertical lines indicate the $D\eta$ and $D_s \bar{K}$ thresholds. Error bands have been obtained by  Monte Carlo  propagating  the uncertainties of the LECs quoted in Ref.~\cite{Liu:2012zya}.}
\label{fig:in3cc}
\end{center}
\end{figure*}
\begin{figure*}[t]
\begin{center}
\includegraphics[width=0.9\textwidth]{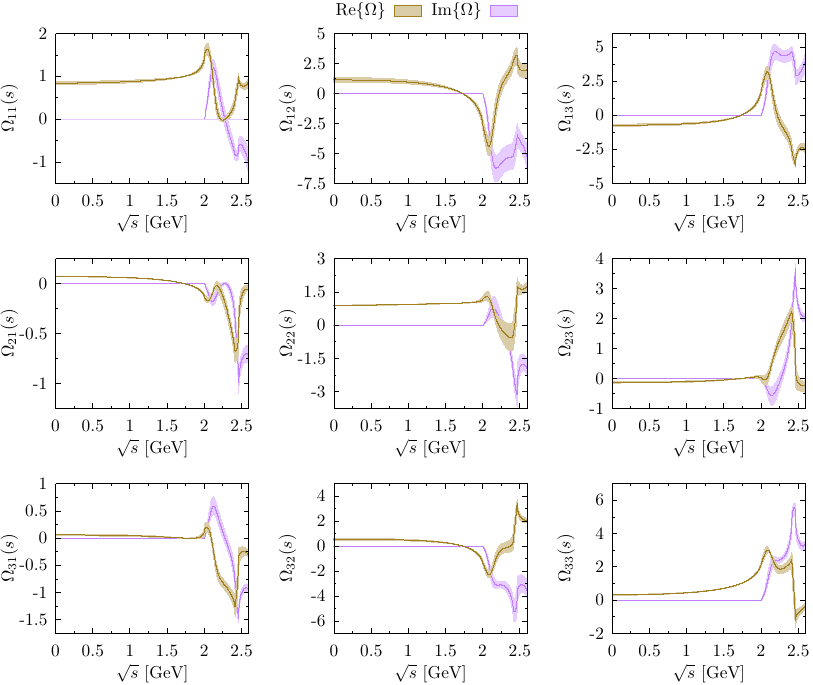}
\caption{Charm $(S,I)=(0,1/2)$  Omn\`es matrix solution of the MO integral equation \eqref{eq:MOpro} with the contour condition $\Omega\left((m_D-M_\pi)^2\right) = \mathbb{I}$, and asymptotic  phase shifts $\delta_{D\pi}(\infty)=3\pi$,  $\delta_{D\eta}(\infty)=0$ and  $\delta_{D_s\bar K}(\infty)=0$. Error bands have been obtained by  Monte Carlo  propagating  the uncertainties of the LECs quoted in Ref.~\cite{Liu:2012zya}.\label{fig:out3cc}}
\end{center}
\end{figure*}
\paragraph{\underline{Charm sector}:} 
Here we consider three channels, $D\pi$ ($1$), $D\eta$ ($2$), and $D_s \bar{K}$ ($3$), and above all thresholds, the $S$-matrix can be still specified\footnote{The $T$-matrix is obtained from 
Eq.~\eqref{eq:defS3channels}.} by the elastic parameters, i.e., three phase shifts and three inelasticities~\cite{Waldenstrom:1974zc,Lesniak:1996qx},
\bea
\label{eq:defS3channels}
S(s)=
\left(
\begin{array}{ccc}
\eta_1e^{2i\delta_1}& \gamma_{12}e^{i\phi_{12}}& \gamma_{13}e^{i\phi_{13}}\\
 \gamma_{12}e^{i\phi_{12}}&\eta_2e^{2i\delta_2}& \gamma_{23}e^{i\phi_{23}}\\
 \gamma_{13}e^{i\phi_{13}}& \gamma_{23}e^{i\phi_{23}}&\eta_3e^{2i\delta_3}\\
\end{array}
\right) ,
\eea
Furthermore, the parameters in the off-diagonal elements are related to the diagonal ones
$\delta_i$ and $\eta_i$ by
\bea
\label{eq:gammas} 
\gamma_{ij}^2&=&\frac{1}{2}\left(1+\eta_k^2-\eta_i^2-\eta_j^2\right), \quad  i\neq
j \neq k \neq i \, ,\nonumber\\
\phi_{ij}&=&\delta_i+\delta_j+\alpha_{ij}+{\rm mod}(\pi),\quad  i,j,k= 1,2,3 \,,\nonumber
\eea
and $\alpha_{ij}$ is determined as
\bea
\sin\alpha_{ij}=\sqrt{\frac{1}{4\eta_i\eta_j}\bigg[\frac{\gamma_{ik}^2\gamma_{jk}^2}{\gamma_{ij}^2}-(\eta_i-\eta_j)^2\bigg]}\equiv X_{ij}\ .
\eea
Note that the solutions for $\alpha_{ij}$ can be either ${\rm \arcsin}(X_{ij})$ 
or $\pi-{\rm \arcsin}(X_{ij})$. The inelasticity parameters should satisfy the following boundary conditions:
\begin{eqnarray}
 0 &\le & \eta_i \le 1 \,, \nonumber \\
 |1-\eta_j-\eta_k| &\le& \eta_i \le 1-|\eta_j-\eta_k|, \quad  i\neq
j \neq k \, .
\end{eqnarray}
To solve the MO integral Eq.~\eqref{eq:MOpro}, we use a $T$-matrix  similar to that in Eq.~\eqref{eq:in2cc}, with the obvious substitution of $T_{U}^{(1,0)}(s)$ by $T_{U}^{(0,\frac{1}{2})}(s)$. 
In addition, $s_{\rm th}= (m_D+M_\pi)^2$ and we now  take $s_m = (2.6~{\rm GeV})^2$. Phase shifts, inelasticities and amplitude moduli from  $T_{U}^{(1,0)}$ are displayed in Fig.~\ref{fig:in3cc} for $s_{\rm th}\le s \le s_m$. Above $s_m = (2.6~{\rm GeV})^2$, $T_H$ is constructed from Eq.~\eqref{eq:defS3channels} using  interpolating parameterizations for phase shifts and inelasticities similar to those given in Eq.~(\ref{eq:in2asy}),  imposing continuity of phase shifts and of the $T$-matrix, and taking
\bea
\delta_1(\infty)&=&3\pi\ ,\quad \delta_{i}(\infty)=0,\quad i=2,3 \,,\nonumber\\
\eta_j(\infty)&=&1\ ,\quad j=1,2,3\ .\label{eq:dpi.eigenps}
\eea
With the inputs specified above, the three-dimensional $(S,I)=(0,1/2)$  Omn\`es matrix can be numerically computed  and its complex elements are shown in
Fig.~\ref{fig:out3cc} up to $\sqrt{s}\le 2.6$ GeV.

\begin{figure*}[t]
\begin{center}
\includegraphics[width=0.89\textwidth]{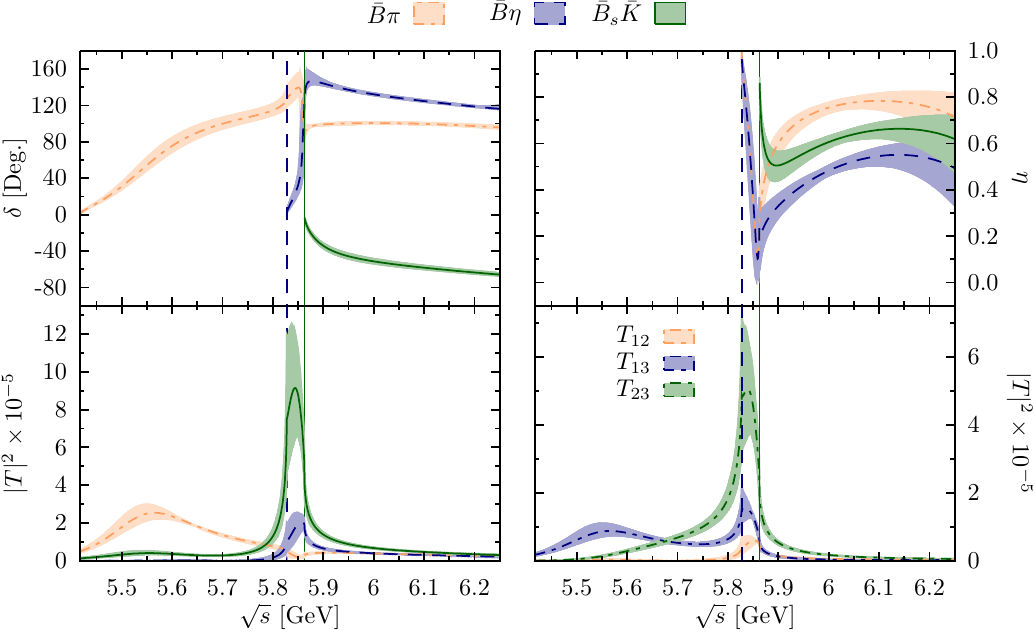}
\caption{Phase shifts, inelasticities  and amplitude moduli from $T_{U}^{(0,1/2)}$ in the bottom sector. The vertical lines indicate the $\bar B\eta$ and $\bar B_s \bar{K}$ thresholds. The values of the
involved LECs in the $\bar{B}\phi$ interactions are determined from their
analogues in the charm sector by imposing the heavy-quark mass scaling rules  discussed in the Appendix~\ref{sec:scaling.h}. Error bands have been obtained by  Monte Carlo  propagating  the uncertainties of the LECs quoted in Ref.~\cite{Liu:2012zya}.}
\label{fig:in3ccBott}
\end{center}
\end{figure*}

\begin{figure*}[t]
\begin{center}
\includegraphics[width=0.9\textwidth]{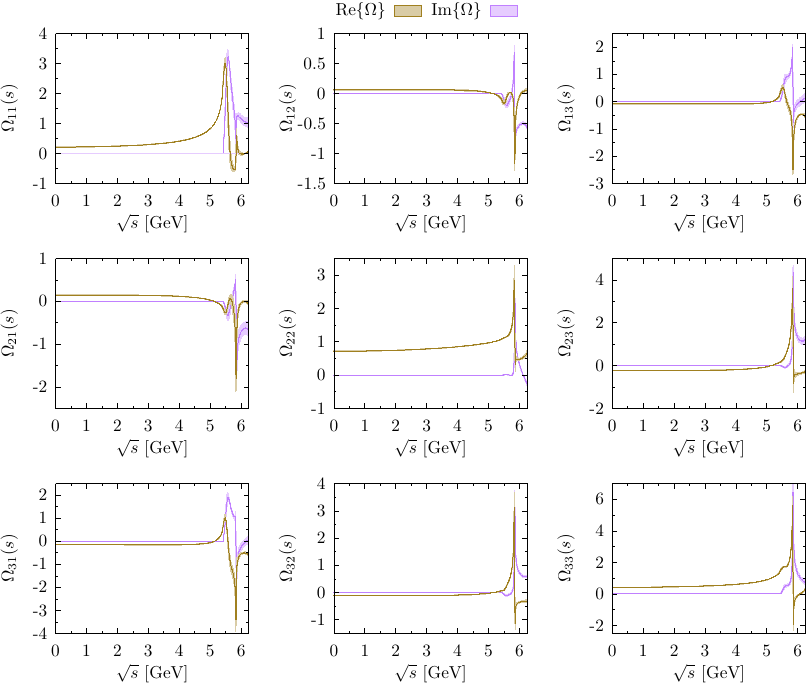}
\caption{Bottom $(S,I)=(0,1/2)$  Omn\`es matrix solution of the MO integral equation \eqref{eq:MOpro} with the contour condition  $\Omega\left((m_B-M_\pi)^2\right) = \mathbb{I}$, and asymptotic  phase shifts $\delta_{\bar B\pi}(\infty)=2\pi$,  $\delta_{\bar B\eta}(\infty)=2\pi$ and  $\delta_{\bar B_s\bar K}(\infty)=0$.  Error bands have been obtained by  Monte Carlo  propagating  the uncertainties of the LECs quoted in Ref.~\cite{Liu:2012zya}.\label{fig:out3ccBott}}
\end{center}
\end{figure*}

\paragraph{\underline{Bottom sector}: \label{sec:ext-bottom}}  In Figs.~\ref{fig:in3ccBott} and \ref{fig:out3ccBott}, we show phase shifts, inelasticities  and the  solution of the 
MO integral equation  for  the $(S,I)=(0,1/2)$ channel in the bottom sector. The chiral amplitudes\footnote{The values of the
involved LECs in the $\bar{B}\phi$ interactions are determined from their
analogues in the charm sector by imposing the heavy-quark mass scaling rules \cite{Albaladejo:2016lbb} discussed in the Appendix~\ref{sec:scaling.h}.} are used in  Eq.~\eqref{eq:OmnesUDR} up to $s_m=(6.25~{\rm GeV})^2$, and from there on, the asymptotic forms of the amplitudes  are employed. As we will show in the next section, in the case of $\bar B$ decays, the accessible phase space is quite large, and  $q^2$ varies from around $m_B^2$ at zero recoil [$q^2_{\rm max} = (m_B-M_\phi)^2$] down to zero, when the energy of the outgoing light meson is about $m_B/2$ far from the chiral domain. The $\bar B \pi$ scalar form factor decreases by a factor of five, and the LQCD results around $q_{\rm max}^2$ and the LCSR predictions in the vicinity of $q^2=0$ are not linearly connected. In the present approach, as we will discuss, we multiply the MO matrix $\Omega$ by a rank-one polynomial, and thus the extra curvature  provided by the  MO matrix becomes essential. While $\Omega(s)$ around $q^2_{\rm max}$  is rather 
insensitive to the adopted asymptotic behaviour of the 
$T$-matrix, since it is dominated by the integration region close to 
threshold ($s< s_m$) where the chiral amplitudes are being used\footnote{In general, the MO matrix in the chiral domain,  between the $q^2_{\rm max}$  and scattering (below $s_m$) regions  is rather insensitive to the high energy behaviour of the amplitudes.}, this is not the case for low values of $q^2$ close to  $0$, quite far from the two-body scattering thresholds. 
This unwanted dependence, due to the large extrapolation,  could be compensated in the form factors by using higher rank polynomials, but that would introduce additional undetermined  parameters. Conversely, this dependence of $\Omega(0)$, relative to the results at $q^2_{\rm max}$,  on the details of the amplitudes at high energies will be diminished by solving a MO integral equation involving several subtractions, instead of the unsubtracted one of Eq.~\eqref{eq:OmnesUDR}. This, however, will also introduce some more free parameters~\cite{Flynn:2007ki}. The situation is better in the charm sector, where the needed $q^2-$range is much reduced, and thus most of the contributions to the MO matrix come from integration region within the chiral regime.
Indeed in the bottom sector we need to use $\delta_1(\infty)=2\pi, \delta_2(\infty)=2\pi$ and $\delta_3(\infty)=0$, instead of the choice of Eq.~\eqref{eq:dpi.eigenps} used for the charm decays, to find acceptable fits to the LQCD and LCSR predictions of the $\bar B\pi$ and $\bar B_s K$ scalar form factors. With this choice, we find theoretically sound fits where the LECs\footnote{These are $\beta_1^P$ and $\beta_2^P$, to be introduced in Subsec.~\ref{sec:SC}, that appear in the chiral expansion of the form factors at NLO.} that determine the rank-one Omn\`es polynomials describe the LQCD data close to $q^2_{\rm max}$, within  the range of expected validity of the chiral expansion, while the LCSR results are reproduced thanks to the non-linear behaviour encoded in the MO matrix $\Omega(s)$. This picture will be reinforced  by the consistent results that will be obtained, assuming a reasonable effect of the HQFS breaking terms, from combined fits to the $D\to \pi/\bar{K}$ and the  $\bar{B}\to\pi$ and  
$\bar{B}_s\to K$ scalar form 
factors. 

We do not really have an explanation of why the above choice of phase shifts at infinity works better in the bottom sector than the usual one in Eq.~\eqref{eq:dpi.eigenps} and adopted in the charm meson decays. We would like, however, to mention the different behaviour of the unitarized chiral phase shifts in the charm and bottom sectors. In both cases, the chiral amplitudes give rise to two resonances~\cite{Albaladejo:2016lbb}: the first one, the non-strangeness flavor partner of the $D_{s0}^\ast(2317)$, quite broad, and located around 100 MeV above the $D\pi$ or $\bar B \pi$ thresholds and the second one placed below the heaviest of the thresholds, $D_s\bar K$ and $\bar B_s\bar K$, respectively. In the charm sector, the second resonance does not produce clear signatures in the phase shifts of the two open channels $D\pi$  and $D\eta$, while it is clearly visible in the phase shifts of the bottom $\bar B\pi$  and $\bar B\eta$ channels. Moreover, the second resonance is significantly narrower for the latter 
heavy-quark sector than for the former one (70 
MeV versus 270 MeV). 

Note also that now $\lim_{s\to \infty} \sum_{i=1}^3 \delta_i(s) = 4\pi> 3\pi$, which implies a slightly  faster decreasing of the MO matrix elements at high energies. 

\subsection{Chiral expansion of the form factors and the MO polynomial}
\label{sec:SC}
Once the Omn{\`e}s matrix is obtained, the form factor $\vec{\cal F}(s)$ is determined, according to Eq.~(\ref{eq:FFmatrix}), up to a polynomial $\vec{\cal
P}(s)$ that contains unknown coefficients. We will match the dispersive and the NLO chiral representations of the form factors in a region of values of $s$ where the latter are supposed to be still valid. Besides the theoretical benefit of this constraint, it has also the practical advantage of expressing the coefficients of the polynomials in terms of the few LECs used in the chiral expansion of the form factors. Since, as will be discussed below (cf.~Eqs.~\eqref{eq:f+chpt} and \eqref{eq:f0chpt} and the discussion that follows), the NLO chiral expansion of the form factors used here is appropriate only up to terms linear in $s$, we should also take  linear forms for the MO polynomials,
\bea\label{eq:poly}
\vec{\mathcal{P}}(s)=\vec{\alpha}_0+\vec{\alpha}_1\,s \ .
\eea
Since the Omn\`es matrix elements asymptotically behave as $1/s$ [see Eq.~\eqref{eq:assydet}], due to the chosen asymptotic conditions, this implies that the form factors will tend to a constant\footnote{This not strictly true in the case of $\bar B_{(s)}-$decays since, as discussed above, different  asymptotic conditions have been assumed in the bottom sector and  the Omn\`es matrix elements are expected to decrease slightly faster than  $1/s$. } for $s\to \infty$. Note that one would rather expect the form factors to vanish in this limit \cite{Lepage:1980fj}. To achieve such asymptotic behaviour one should employ order zero polynomials. However, since we are interested in the region $0 \leqslant s \leqslant s_{\rm max}$,  with $s_{\rm max}$ in the vicinity of  $(m_H-M_\phi)^2$, we prefer to keep the linear behaviour of the polynomials, since this allows for 
a better matching  of the coefficients $\vec{\alpha}_{0,1}$ with the LECs that appear in the NLO chiral calculation of the form factors.

\subsubsection{Form factors in heavy meson chiral perturbation theory}

\begin{figure}[t]
\begin{center}
\includegraphics[scale=0.8]{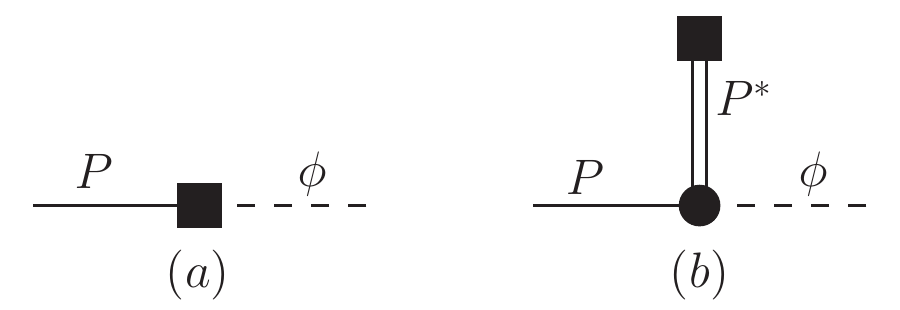}
\caption{Topologies of the relevant Feynman diagrams contributing to the hadronic matrix elements.  The solid circle denotes the LO ${\cal P}{\cal P}^\ast\phi$ interaction and the solid square represents the left-handed current.\label{fig:LO}}
\end{center}
\end{figure}

The leading-order (LO) coupling of the charm ($D$ and $D_s$) or bottom ($\bar B$ and $\bar B_s$) mesons  to the Nambu--Goldstone bosons of the spontaneous
breaking of the approximate chiral symmetry of QCD, through the charged-current left-handed
current $J^\mu = \left(\bar{Q}\gamma^\mu_L u,\bar{Q}\gamma^\mu_L d, \bar{Q}\gamma^\mu_L s \right)^T$, with $Q=c,b, $ and  $\gamma^\mu_L = \gamma^\mu(1-\gamma_5)$,  is described by the following chiral effective Lagrangian~\cite{Wise:1992hn,Burdman:1992gh,Yan:1992gz} 
\bea
\label{eq:lagLO} 
\mathcal{L}_0=\sqrt{2}f_{\mathcal P}\left(i\, \mathring{m}{ \cal P}^\ast_\mu +\partial_\mu 
{\cal P}\right) u^\dagger J^\mu \ .
\eea
where  ${\cal P}$ and ${\cal P}^\ast$  are the pseudoscalar and vector heavy-light mesons with content $(Q\bar u,Q \bar d,Q \bar s)$, 
respectively, which  behave as $SU(3)$ light flavor triplets. Here $\mathring{m}$ denotes the degenerate mass of the $P_{(s)}$ and $P^\ast_{(s)}$ mesons in the chiral and heavy-quark 
limits, and $f_{\cal P}$ is the pseudoscalar heavy-light meson decay constant
defined as
\begin{equation}
\langle0|J^\mu|{\cal P}(p_1)\rangle=i\,\sqrt{2}f_{\cal P} \,p_1^\mu\ .
\end{equation}
The chiral block is defined by $u^2=U=\exp[i\sqrt{2}\Phi/F_0]$, where $\Phi$ is the 
 octet of the Nambu--Goldstone bosons
 \begin{equation}
\Phi=\begin{pmatrix}
   \frac{1}{\sqrt{2}}\pi^0 +\frac{1}{\sqrt{6}}\eta  & \pi^+ &K^+  \\
     \pi^- &  -\frac{1}{\sqrt{2}}\pi^0 +\frac{1}{\sqrt{6}}\eta&K^0 \\
     K^-&\bar{K}^0&-\frac{2}{\sqrt{6}}\eta
\end{pmatrix} ,
\end{equation}
with $F_0$ the pion decay constant in the chiral limit (we will take the physical value for the decay constant $F_0\simeq92 $~MeV). The relevant NLO chiral effective
 Lagrangian reads~\cite{Burdman:1992gh}
\begin{equation}
{\cal L}_1=-\beta_1^P{\cal P}\,u\,(\partial_\mu U^\dagger) J^\mu -\beta_2^P(\partial_\mu\partial_\nu {\cal P})\,u\,(\partial^\nu U^\dagger)\,J^\mu\ .
\end{equation}
We need the LO $\mathcal{P}\mathcal{P}^\ast\phi$ interaction as well, which is given by~\cite{Wise:1992hn,Burdman:1992gh,Yan:1992gz} ,
\begin{equation}\label{eq:lagppstp} {\cal
L}_{\mathcal{P}\mathcal{P}^\ast\phi}=\tilde{g}\left(\mathcal{P}^\ast_{\mu}u^\mu \mathcal{P}^\dagger+\mathcal{P}\,u^\mu \mathcal{P}^{\ast\dagger}_\mu\right)\ ,
\end{equation}
where $u_\mu=i(u^\dagger\partial_\mu u-u\,\partial_\mu u^\dagger)$ and $\tilde{g}\sim g \mathring{m}$, with $g\sim 0.6$ a dimensionless and heavy quark mass independent constant. The topologies of relevant Feynman diagrams are shown in Fig.~\ref{fig:LO}.  The vector and scalar form factors, in the (strangeness, isospin) basis,  at ${\cal O}(E_\phi)$ (i.e., NLO) in the chiral expansion read~\cite{Burdman:1992gh}
\begin{align}
f_{+}^{[P\phi]^{(S,I)}}(s)= & \frac{\mathcal{C}_{[P{\phi}]}^{(S,I)}}
{\sqrt{2}F_0}\bigg[\frac{f_P}{\sqrt{2}}+\sqrt{2}\frac{\tilde{g}\,\mathring{m}\,f_P}{m_R^{2}-s}
 \nonumber\\
& + \beta_1^P-\frac{\beta_2^P}{2}(\Sigma_{P\phi}-s) \bigg],\label{eq:f+chpt}\\
f_0^{[P\phi]^{(S,I)}}(s) = & \frac{\mathcal{C}_{[P{\phi}]}^{(S,I)}}
{\sqrt{2}F_0}\bigg[\bigg(\sqrt{2}\frac{\tilde{g}\,\mathring{m}f_P}{m_R^{2}}+\beta_1^P\bigg)
\frac{\Delta_{P\phi}-s}{\Delta_{P\phi}}
 \nonumber\\ 
& +\bigg(\sqrt{2}f_P-{\beta_2^P}(\Sigma_{P\phi}-s)\bigg)\frac{\Delta_{P\phi}+s}{2\,\Delta_{P\phi}}
\bigg],\label{eq:f0chpt}
\end{align}
where $\Delta_{P\phi}=m_P^2-M_\phi^2$ and $\Sigma_{P\phi}=m_P^2+M_\phi^2$ with 
$P\in\{Q\bar s, Q\bar d, Q\bar u\}$ and $\phi\in\{\pi,~K,\bar K~\eta\}$.  The
coefficients $\mathcal{C}_{[P{\phi}]}^{(S,I)}$ are collected in 
Table~\ref{TabCoe}. Moreover, we have fixed $m_R$ to  $m_{D^\ast}$ ($m_{B^\ast}$) and to $m_{D^\ast_s}$ ($m_{B^\ast_s}$) for the 
$(S,I)=(0,{1}/{2})$ and $(S,I)=(1,0)$ charm (bottom) channels, respectively. In principle, at LO in the heavy quark expansion, $m_R$ should be set to $\mathring{m}$,  however the use of the physical vector mass is quite relevant for the vector form factor, because of the propagator structure, though it has much less relevance for the scalar form factor that we study in this work. We should also note that all kinematical factors are always calculated using physical masses of the involved mesons. It is worthwhile to notice that $s$ is of the order $m_P^2 \sim \mathcal{O}(E_\phi^0)$, $\left(\Delta_{P\phi}-s\right)\sim \mathcal{O}(E_\phi)$ and $\left(\Sigma_{P\phi}-s\right)\sim \mathcal{O}(E_\phi)$, and $\left(\Delta_{P\phi}+s\right) \sim \mathcal{O}(E_\phi^0)$, so that a small change of $\mathcal{O}(E_\phi)$ in $\left(\Delta_{P\phi}+s\right)$ only leads to a higher order effect. Thus, $\left(\Delta_{P\phi}+s\right)$ should be regarded as basically a constant with $s\sim m_P^2$, and the expression 
in Eq.~\eqref{eq:f0chpt} should be matched to a rank-1 MO polynomial as mentioned before (see below for details of the matching).  

Finally, we should mention that the LECs  $\beta_1^P$ and  $\beta_2^P$  scale with the heavy quark mass as~\cite{Burdman:1992gh}
\begin{equation}
 \beta_1^P \sim \sqrt{m_P},\qquad \beta_2^P \sim 1/\sqrt{m_P^3} \label{eq:b1b2-scal}
\end{equation}
neglecting logarithmic corrections. 

\begin{table}[t]
\caption{Strangeness-isospin coefficients appearing in the chiral expansion of the form factors.}\label{TabCoe}
\begin{tabular}{c|cc|ccc}
\hline\hline
$(S,I)$       & \multicolumn{2}{c|}{$(1,0)$} & \multicolumn{3}{c}{$(0,\frac{1}{2})$} \\
{\rm channel} &$DK$ & $D_s\eta$              & $D\pi$ & $D\eta$   & $D_s\bar{K}$\\ 
& &             & $\bar B\pi$ & $\bar B \eta$   & $\bar B_s\bar{K}$\\ 
\hline
$\mathcal{C}$ & $-\sqrt{2}$ &$ -\sqrt{\frac{2}{3}}$ & $\sqrt{\frac32}$ & $\frac{1}{\sqrt{6}}$ &  1\\
\hline\hline
\end{tabular}
\end{table}

\subsubsection{Matching}\label{subsec:matching}

At energies close to the thresholds, the scalar form factors in Eq.~\eqref{eq:FFmatrix} should have
the same structure as the ones obtained from chiral perturbation theory,
given in Eq.~\eqref{eq:f0chpt}. We match the two representations at a point
$s=s_0$ located in the valid region of the chiral expansion. Namely, we take $s_0 = q^2_\text{max} = (m_P - M_\pi)^2$ for the $(0,1/2)$ form factors, and $s_0 = (m_D - M_K)^2$ for the charm $(1,0)$ case, since this is the point in which the momentum of the lightest meson is zero. Imposing that the dispersive form factors and their first derivative to be equal to the chiral ones at $s=s_0$, the
coefficients in the polynomials, Eq.~\eqref{eq:poly}, can be expressed as:
\begin{align}\label{eq:polycoeffmatch}
\vec{\alpha}_0 & = \Omega^{-1}(s_0)\cdot\vec{\mathcal{F}}_{\chi}(s_0)-\vec{\alpha}_1\,s_0,\\
\vec{\alpha}_1 & = \Omega^{-1}(s_0)\cdot\left[\vec{\mathcal{F}}^\prime_{\chi}(s_0)-\Omega^\prime(s_0)\cdot\Omega^{-1}(s_0)\cdot\vec{\mathcal{F}}_{\chi}(s_0)\right],\nonumber
\end{align}
where the $^\prime$ stands for a derivative with respect to $s$. The vectors $\vec{\cal{F}}_\chi$ contain the chiral form factors,
\begin{align}
\vec{{\cal F}}_\chi^{(0,{1\over2})}(s) & \equiv\left(f_0^{{D\pi}^{(0,\frac{1}{2})}},f_0^{{D\eta}^{(0,\frac{1}{2})}},f_0^{{D_sK}^{(0,\frac{1}{2})}}\right)^T~,\\
 \vec{{\cal F}}_\chi^{(1,0)}(s) & \equiv\left(f_0^{{DK}^{(1,0)}},f_0^{{D_s\eta}^{(1,0)}}\right)^T~,
\end{align}
with all the elements given in Eq.~\eqref{eq:f0chpt}. Similar expressions are used for the $(0,{1\over2})$ channel in the bottom sector. In other words, the vectors $\vec{\mathcal F}_\chi(s)$ contain the form factors defined in Eqs.~\eqref{eq:mo.dpi} and \eqref{eq:mo.dk}, but computed according to the chiral expansion.

It is worth noting that the NLO LECs $\beta_1^P$ and $\beta_2^P$ determined from a fit to data using the MO scheme would have some residual dependence on the matching point. To minimize such dependence, we have 
chosen $s_0=q_{\rm max}^2$, where the momentum of the Goldstone bosons is close to zero and higher order chiral corrections are expected to be small. Different choices of the matching point, within  the chiral regime, will amount to changes in the fitted (effective)  $\beta_1^P$ and $\beta_2^P$ LECs driven by higher order effects. 

In the charm $(1,0)$ sector, due to the presence of the $D^\ast_{s0}(2317)$ state as a bound state in the $T$-matrix, the solution of Eq.~\eqref{eq:FFmatrix} gets modified. The contribution from $D_{s0}^\ast(2317)$ is easily incorporated as follows:
\begin{equation}\label{eq:Poly2317}
\Omega^{(1,0)}\cdot\vec{\cal P}^{(1,0)}(s)\to\Omega^{(1,0)}\cdot\bigg\{\frac{\beta_0\,\vec{\Gamma}}{s-s_p}+\vec{\cal P}^{(1,0)}(s)\bigg\},
\end{equation}
where $\beta_0$ is an unknown parameter which characterizes the coupling of
$D_{s0}^\ast(2317)$ to the left-hand current. $\vec{\Gamma}$ contains the
couplings of the $D_{s0}^\ast(2317)$ to the $DK$-$D_s\eta$ system\footnote{Note that the first term in the bracket of Eq.~\eqref{eq:Poly2317} should have a more general form, $\frac{\vec{\beta}_0}{s-s_p}$, with $\vec{\beta}_0$ a vector with two independent components, $\vec{\beta}_0 = (\beta_0^a, \beta_0^b)^T$. The specific form in Eq.~\eqref{eq:Poly2317} reduces the number of free parameters, by forcing $\beta_0^a/\beta_0^b = g_{DK}/g_{D_s \eta}$. On the other hand, this has the effect that the form factors $f_0^{DK}$ and $f_0^{D_s \eta}$ are not exactly independent of the choice of the point $s_n$ where one normalizes the Omn\`es matrix, $\Omega(s_n) = \mathbb{I}$. Nonetheless, we have checked that this choice, varying $s_n$  from zero  to $q^2_\text{max}$,  has no practical effect in the determination of $\beta_0$, which indicates that our assumption is reasonable. We also remark that this discussion has no effect at all in the $(0,1/2)$ sector.}, namely,
$\vec{\Gamma}=(g_{DK},g_{D_s\eta})^T$. This bound state is dynamically generated in the unitarized amplitudes 
given in Ref.~\cite{Liu:2012zya}, that we employ here. The couplings are computed from the residue of the amplitude at the pole,
\begin{equation}
T_{ij}(s) = \frac{g_i g_j}{s-s_p} + \ldots
\end{equation}
The $D_{s0}^\ast(2317)$ pole position $s_p$, together with $g_{DK}$ and $g_{D_s\eta}$, are collected in
Table~\ref{tab:pole}.

\begin{table}[t]
\caption{Properties of the $D_{s0}^\ast(2317)$ pole from the unitarized chiral amplitudes derived in Ref.~\cite{Liu:2012zya}.\label{tab:pole}}
\centering
\begin{tabular}{ccc}
\hline
$\sqrt{s_{p}}$~[MeV]&$g_{DK}$~[GeV]& $g_{D_s\eta}$~[GeV]\\
\hline
$2315^{+18}_{-28}$ 			& $9.5^{+1.2}_{-1.1}$	 	& $7.5^{+0.5}_{-0.5}$ \\
\hline
\end{tabular}
\end{table}

\section{Numerical results and discussion\label{sec:NR}}
\begin{table}[t]
\caption{Masses and decay constants (in MeV units) used in this work and  taken from the PDG~\cite{Olive:2016xmw}  and FLAG~\cite{Aoki:2016frl} reviews.\label{tab:inputmass}} 
\centering
\begin{tabular}{cc|cc|cc}\hline\hline
\multicolumn{2}{c}{\text{Goldstone}}  & \multicolumn{2}{c}{\text{charm sector}} &\multicolumn{2}{c}{\text{bottom sector}} \\
\hline
$M_\pi$  & $139$  & $m_D$          & $1869.6$ & $m_B$          & $5279.5$ \\
$M_K$    & $496$  & $m_{D_s}$      & $1968.5$ & $m_{B_s}$      & $5366.8$ \\
$M_\eta$ & $547$  & $f_D$          & $147.6$  & $f_B$          & $135.8$  \\
$F_0$    & $92.4$ & $f_{D_s}$      & $174.2$  & $f_{B_s}$      & $161.5$ \\
         &        & $m_{D^\ast}$   & $2008.6$ & $m_{B^\ast}$   & $5324.7$ \\
 &       & $m_{D_s^\ast}$ & $2112.1$ &  &  \\
\hline\hline
\end{tabular}
\end{table}
So far, the theoretical MO representations of the scalar form factors have been
constructed. In this section, we want to confront the so-obtained form factors
to the LQCD and LCSR results.   In what follows, we first fit to the $\bar B \to \pi $ and $\bar B_s \to \bar K$ scalar form factors, where we expect the $1/m_P$ corrections  to the chiral expansion in Eq.~\eqref{eq:f0chpt} to be substantially suppressed.  Next, we will carry out a combined fit to all the data in both charm and bottom sectors, by adopting some (approximate) heavy-quark flavor scaling rule~\cite{Burdman:1992gh} for the $\beta_1^P $ and $\beta_2^P$ LECs in Eq.~\eqref{eq:f0chpt}.
Using  the results of our combined fit, we will: i) determine the CKM elements, $|V_{cd}|$, $|V_{cs}|$ and
$|V_{ub}|$,  ii) predict form factors,  not computed  in LQCD yet, and 
that can be used to over-constrain the CKM matrix elements from analyses involving more  semileptonic decays, and iii) predict the different form factors above the $q^2-$regions accessible in the semileptonic decays, up to moderate energies amenable to be described using the 
unitarized coupled-channel chiral approach.

Masses and decay constants used in this work are compiled
in Table~\ref{tab:inputmass}. In addition, the mass of the heavy-light mesons in the chiral limit, see
Eq.~(\ref{eq:lagLO}), is set to $\mathring{m}=(m_P+m_{P_s})/2$, for simplicity the same average is used to define $\bar m_P$ in the Appendix~\ref{sec:scaling.h} and in the relations given in Eq.~\eqref{eq:scaling.beta}. The $\mathcal{P}\mathcal{P}^\ast\phi$ axial
coupling constant $\tilde{g}$ in  Eq.~(\ref{eq:lagppstp}) can be fixed by calculating 
the decay width of $D^{\ast+}\to D^0\pi^+$~\cite{Olive:2016xmw},
which leads\footnote{Errors on $g$ determined  from the decay $D^{\ast+}\to D^0\pi^+$ are very small of the order of 1\%.  } to $g\sim 0.58$ and hence $\tilde{g}_{D^*D\pi}\sim 1.113$ GeV. In the bottom sector we use a different value for $g$, around 15\% smaller, consistent with the lattice calculation of Ref.~\cite{Ohki:2008py}, where  $g\sim 0.51$ 
(or $\tilde{g}_{\bar B^*\bar B\pi}\sim 2.720$ GeV) was found. Note that the difference is consistent with the expected size of  heavy-quark-flavor symmetry violations. In addition, there exist sizable  $SU(3)$ corrections 
to the overall size of the ${\cal P}^*$ pole contribution to both $f_+$ and $f_0$ form factors. Thus, such contribution is around $\sim 20\%$ smaller for $\bar B^* \bar B_s K $ than for $\bar B^*\bar B\pi$ ~\cite{Albertus:2005ud, Albertus:2014gba}. According to \cite{Colangelo:2002dg} this suppression is mainly due to a factor $F_\pi/F_K\sim 0.83$~\cite{Aoki:2016frl}. We will implement this correction in the pole contribution to $f_0$ in  
Eq.~\eqref{eq:f0chpt} when the Goldstone boson is  either a kaon or an eta meson (for simplicity, we also take $F_\eta \approx F_K$), and both in the bottom and charm sectors. 

\subsection{Fit to the LQCD+LCSR results in the bottom sector\label{sec:fit}}

We are first interested
in the $\bar B\to\phi$ transitions induced by the $b\to u$ flavour-changing
current, which include $\bar{B}\to\pi$, $\bar{B}\to \eta$ and $\bar{B}_s\to K$.
The scalar form factors involved in those transitions can be related to
the Omn\`es matrix through 
\bea\label{eq:mo.bpi}
\left(
\begin{array}{c}
  \sqrt{\frac{3}{2}}  f_0^{\bar{B}^0\to \pi^+}(s) \\
f_0^{B^-\to{\eta}}(s)  \\
f_0^{\bar{B}_s^0\to{K^+}}(s)
\end{array}
\right)=
\Omega^{(0,{1\over2})}_{\bar{B}}(s)\cdot \vec{\cal P}^{(0,{1\over2})}_{\bar{B}}(s)\ .
\eea
Within the present approach, and considering just rank-one MO polynomials, there are only two undetermined parameters: the NLO LECs $\beta_1^P$ and $\beta_2^P$ that appear in the chiral expansion of the form factors in Eq.~\eqref{eq:f0chpt}. We fit these parameters, in the bottom sector,  to  LQCD   (UKQCD~\cite{Flynn:2015mha}, HPQCD~\cite{Dalgic:2006dt, Bouchard:2014ypa} and Fermilab Lattice \& MILC (to be referred to as FL-MILC for brevity)~\cite{Lattice:2015tia})  and  LCSR~\cite{Duplancic:2008ix, Duplancic:2008tk} results for the scalar form factors in $\bar{B}^0\to \pi^+$ and $\bar{B}_s^0\to{K^+}$ semileptonic decays. Lattice results are not available for the whole kinematic region accessible in the decays, and they are restricted to large values of $q^2\ge 17$ GeV$^2$, where momentum-dependent discretization and statistical errors are under control. To constrain the behaviour of the
scalar form factors at small values of $q^2$, we take four LCSR points (equally-spaced) in the interval $q^2=0-6$ GeV$^2$ for each decay.  

The UKQCD Collaboration~\cite{Flynn:2015mha} provides data for both $\bar B\to\pi$ and $\bar B_s\to K$ form factors together with statistical and systematic  correlation matrices for a set of three form factors computed at different $q^2$ (Tables VIII and IX of this reference). In the case of  HPQCD~\cite{Bouchard:2014ypa} $\bar B_s\to K$ and FL-MILC~\cite{Lattice:2015tia}  $B\to\pi$ form factors, we have read off four points from the final extrapolated results (bands) given in these references, since in both cases, originally  only four momentum configurations ($0\to 0$, $0\to 1$, $0\to \sqrt{2}$ and $0 \to \sqrt{3}$) were simulated.  Finally, we also include in the fit the five $B\to\pi$ points provided by the HPQCD Collaboration in the erratum of Ref.~\cite{Dalgic:2006dt}.

Thus, the $\chi^2$ function  reads
\bea
\chi^2&=&(\chi_{\rm cov}^2)^{\bar{B}\to\pi}_{\rm UKQCD}+(\chi_{\rm cov}^2)^{\bar{B}_s\to K}_{\rm UKQCD}+(\chi^2)^{\bar{B}\to\pi}_{\rm FL-MILC}
\nonumber\\
&+&(\chi^2)^{\bar{B}\to\pi}_{\rm HPQCD}+(\chi^2)^{\bar{B}_s\to K}_{\rm HPQCD}\nonumber\\
&+&(\chi^2)_{\rm LCSR}^{\bar{B}\to\pi} + (\chi^2)_{\rm LCSR}^{\bar{B}_s\to K}\ , \label{eq:chi2bottom}
\eea
where $\chi^2$ is the usual uncorrelated Gaussian merit function, and $\chi_{\rm cov}^2$ is defined as
\bea\label{eq:chisq.cov}
\chi^2_{\rm cov}=\sum_{i,j=1}\big[f_0(q^2_i)-f_{0}^i\big]
(\mathcal{C}^{-1})_{ij}\big[f_0(q^2_j)-f_{0}^j\big]\ .
\eea
with the covariance matrix $\mathcal{C}$ constructed out the statistical and systematic correlation matrices and uncertainties given in Ref.~\cite{Flynn:2015mha}. Here $f_0(q^2)$ stands for the theoretical scalar form factor obtained from  the MO representation.

\begin{figure*}[tbh]
\begin{center}
\includegraphics[width=0.9\textwidth]{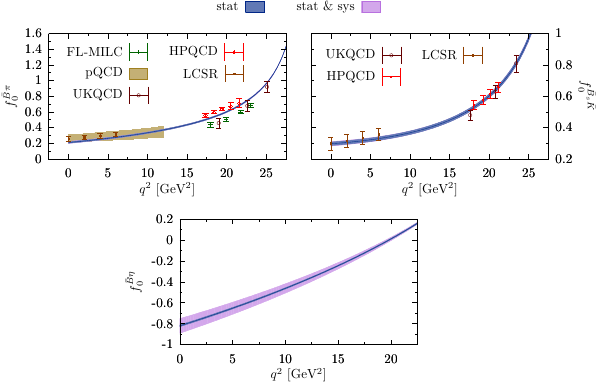}
\caption{Fitted $\bar B^0 \to \pi^+$, $\bar B^0_s \to K^+$ (top) and predicted $\bar B^- \to \eta$ (bottom) scalar form factors. Besides the fitted data (UKQCD~\cite{Flynn:2015mha}, HPQCD~\cite{Dalgic:2006dt, Bouchard:2014ypa},  FL-MILC~\cite{Lattice:2015tia}  and  LCSR~\cite{Duplancic:2008ix, Duplancic:2008tk}),  and for comparison, predictions  from the NLO perturbative QCD approach of Ref.~\cite{Li:2012nk}
for the $\bar B^0 \to \pi^+$ decay are also shown. Statistical (stat) and statistical plus systematic (stat \& sys)  68\%-confident level (CL) bands are also shown. The systematic uncertainties are inherited from  the errors on 
the LECs quoted in Ref.~\cite{Liu:2012zya}, that enter in the definition of the chiral amplitudes, and are added in quadratures to the statistical uncertainties to obtain the outer bands. To estimate the systematic uncertainties for each set of  LECs we re-do the best fit. 
\label{fig:bottomfit}}
\end{center}
\end{figure*}

The chi-squared fit results are
\begin{eqnarray}
 \beta_1^B &=&(0.27\pm 0.12 \pm 0.07)~ {\rm GeV}\nonumber \\
 \beta_2^B &=& (0.037 \pm 0.004\pm 0.003)~ {\rm GeV}^{-1} \label{eq:b1b2besfit}
\end{eqnarray}
with $\chi^2/dof = 4.2$ for 25 degrees of freedom, and a correlation coefficient 0.999
between the two fitted parameters. The first set of errors in the parameters is obtained from the minimization procedure, assuming Gaussian statistics, while the second one accounts
for the uncertainties of the LECs quoted in Ref.~\cite{Liu:2012zya} that enter in the definition of the chiral amplitudes. Such a correlation coefficient so close to 1 indicates that the considered data can not properly disentangle both LECs,\footnote{This can be easily
understood since these LECs enter in the definition of $\vec{\alpha}_0$
and $\vec{\alpha}_1$ in the combinations $\beta_1-m_P^2\beta_2$ and
$\beta_1-m_P(m_P-2M_\phi)\beta_2$ , which are identical up to some 
small $SU(3)$ corrections.} and that different $(\beta_1^B, \beta_2^B)$ pairs belonging to the straight line 
\be
\frac{\beta_1^B-\overline\beta_1^B}{\sigma_{\beta_1^B}}= \frac{\beta_2^B-\overline\beta_2^B}{\sigma_{\beta_2^B}}\, \label{eq:recta}
\ee 
in the vicinity of the best fit values $(\overline \beta_1^B, \overline \beta_2^B)$ quoted in  Eq.~\eqref{eq:b1b2besfit}  lead to similar descriptions of the data (see the dashed-blue line in the right panel of Fig.~\ref{fig:correlac}).  The scalar form factors obtained are displayed in Fig.~\ref{fig:bottomfit}. We find a fair description of the LQCD and LCSR results for the $\bar B^0_s \to K^+$ scalar form factor, while we face some problems for the  $\bar B^0 \to \pi^+$ decay. The large value of $\chi^2/dof$ reported in  Eq.~\eqref{eq:b1b2besfit}  is mainly due to the existing tension between the LQCD results from different collaborations in this latter decay. The disagreement between UKQCD and HPQCD $\bar B^0 \to \pi^+$ scalar form factors was already highlighted  in the top-right panel of Fig. 23 of the UKQCD work~\cite{Flynn:2015mha}, where it is noted that the HPQCD  calculation used only a single lattice spacing.

In addition, as we discussed before,  the $\bar B \pi-$scalar form factor decreases by a factor of five in the $q^2-$range accessible in the decay, and the LQCD results around $q_{\rm max}^2$ and the LCSR predictions in the vicinity of $q^2=0$ are not linearly connected at all. In the current scheme, where only rank-one MO polynomials are being used,  this extra needed curvature should be provided by the $q^2-$dependence of the  MO matrix, $\Omega$, whose behaviour  near $q^2=0$, far from $q^2_{\rm max}$,  is not determined by the behaviour of the amplitudes in the chiral regime. Indeed,  it significantly depends on the high-energy input\footnote{The  results displayed in Fig.~\ref{fig:bottomfit} might suggest that the present approach hardly provides enough freedom to simultaneously accommodate the near $q^2=0$  (LCSR) and $q^2_{\rm max }$ (LQCD) determinations of the $\bar B \pi$ scalar form factor. The situation greatly improves when only the HPQCD, among all LQCD calculations, $\bar B\pi $ results are 
considered in the  $q^2_{\rm max }$ region, being then possible to find an excellent combined description of the LCSR and HPQCD results with $\chi^2=9.65$ for a total of 18 degrees of freedom (see dashed-red curve in the right plot  of Fig.~\ref{fig:correlac}), which leads to $\chi^2/dof= 0.5$. The parameters $\beta_{1,2}^B$ come out still to be almost totally correlated  as in  Eq.~\eqref{eq:b1b2besfit}, and moreover they lie, within great precision, in  the straight line of Eq.~\eqref{eq:recta}, but in the $\beta_1^B\sim 0.7$ GeV region.}.  This is an  unwanted feature, source of systematic uncertainties.  To minimize this problem, in the next subsection we will 
perform a combined fit to 
transitions induced by the $b\to u$ and $c \to d,s$ flavour-changing currents. The latter ones describe $D\to \pi$ and $D\to \bar K$  semileptonic decays for which there exist 
recent and accurate LQCD determinations of the scalar form factors. Moreover, in these latter transitions the $q^2-$ranges accessible in the decays and the  form factor variations  are much limited, becoming thus more relevant the input provided in the chiral regime.

To finish this subsection, we would like to stress that given the large value found for $\chi^2/dof$, statistical errors should be taken with some care. Indeed, one can rather assume some systematic uncertainties affecting our results, that could be estimated by considering in the best fit alternatively only the HPQCD or the UKQCD and the FL-MILC sets of predictions. We will follow this strategy to obtain our final results for the CKM matrix elements and form factors at $q^2=0$ from the combined-fit to charm and bottom decays detailed in the next subsection. 

%

\subsection{Extension to the charm sector and combined fit\label{sec:combinedfit}}

Besides the parameter $\beta_0$ introduced in Eq.~\eqref{eq:Poly2317} to account for the effects on the  $D^\ast_{s0}(2317)$ state in  the $c \to s$ decays, one should also take into account that the LECs  $\beta_1^P$ and $\beta_2^P$ depend on the heavy quark mass. The scaling rules given in Eq.~\eqref{eq:b1b2-scal} can be used to relate the values taken for these LECs in the bottom ($\beta_i^B$) and charm sectors ($\beta_i^D$). We will assume some heavy quark flavor symmetry violations and we will use
\bea \label{eq:scaling.beta}
\frac{\beta_1^D}{\beta_1^B }=\sqrt{\frac{\bar{m}_D}{\bar{m}_B}} (1+\delta), \quad 
\frac{\beta_2^D}{\beta_2^B}=\sqrt{\frac{\bar{m}_B^3}{\bar{m}_D^3}} (1-3\delta)\ 
\eea
where, one should expect the new parameter, $\delta$,  to be of the order  $\Lambda_{\rm
QCD}/\bar m_D$. Note that we are correlating the heavy quark flavor symmetry violations in the LECs $\beta_1$ and $\beta_2$. There is not a good reason for this other than avoiding to include new free parameters.  On the other hand, at the charm scale, one might also expect sizable corrections to the LO prediction $f_0(s) \sim  \mathcal{C}\times f_P /F_0$ of Eq.~\eqref{eq:f0chpt}, even more bearing in mind the large ($40-50\%$) heavy-quark symmetry violations inferred from the  ratio $f_B/f_D$ quoted in Table ~\ref{tab:inputmass}. (Note that  at LO in the inverse of the heavy quark mass, this ratio should scale as $(\bar m_D/\bar m_B)^{1/2}$). Thus, we have also introduced an additional parameter, $\delta^\prime$, defined through  the replacement 
\begin{equation}
 f_P \to f_P \times (1+ \delta^\prime)
\end{equation}
when Eq.~\eqref{eq:f0chpt} is applied to the $c \to d,s$ decays. Thus, we have three new parameters $\beta_0, \delta, \delta^\prime$, which in addition to $\beta_{1,2}^B$,  will be fitted to the LQCD \& LCSR  results for the  scalar form factors in the $\bar B \to \pi$, $\bar B_s \to K$, $D \to \pi$ and $D\to \bar K$ semileptonic decays. 

\begin{figure*}[t]
\begin{center}
\includegraphics[width=0.89\textwidth]{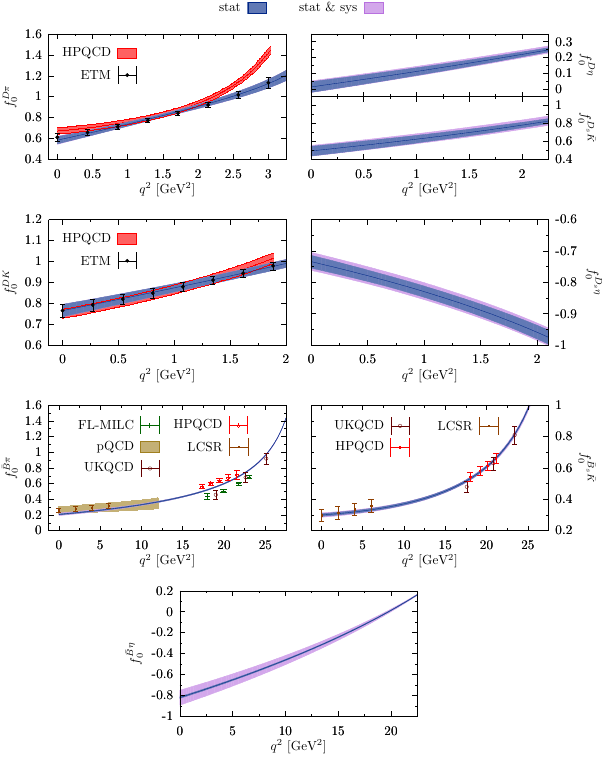}
\caption{Scalar form factors from the  $D_{\ell3}$, $\bar B_{\ell3}$ and $(\bar B_s)_{\ell3}$ combined fit (see Table~\ref{tab:fit.combined} for details). The three bottom panels are similar to those depicted in  Fig.~\ref{fig:bottomfit}, but computed from the results  of the combined best-fit.  The four panels in the first two rows  show form factors for $c\to d, s$ semileptonic transitions. Only ETM results, corrected for some  hypercubic (discretization) effects~\cite{Lubicz:2017syv},  have been considered in the fit of Table~\ref{tab:fit.combined}. For comparison, predictions  from the HPQCD ~\cite{Na:2010uf,Na:2011mc} Collaboration are also displayed.  Differences between ETM and HPQCD sets of $D\to \pi$ and $D\to K$ form factors are clearly visible in the vicinity of $q^2_{\rm max}$, in particular  for the   $D\to \pi$ case.   Statistical (stat) and statistical plus systematic (stat \& sys)  68\%-confident level bands are also given and are calculated as explained in Fig.~\ref{fig:bottomfit}.  Finally, 
predictions for the $D\to \eta $, $D_s\to K $  and $D_s \to \eta$ scalar form factors are also shown. \label{fig:combinedfit}}
\end{center}
\end{figure*}

First we need to incorporate the $c \to d,s$ input into the merit function $\chi^2$, which was defined in Eq.~\eqref{eq:chi2bottom} using only bottom decay results. In the last ten years, LQCD computations of the relevant $D\to\pi$ and $D\to \bar{K}$ semileptonic decay matrix elements have been carried out  by the HPQCD~\cite{Na:2010uf,Na:2011mc} and
very recently by the ETM~\cite{Lubicz:2017syv} Collaborations.  Compared with the
former, the latter corrects for  some hypercubic effects, coming from discretization of a quantum field theory on a 
lattice with hypercubic symmetry~\cite{deSoto:2007ht}, and uses a large sample of kinematics, not restricted in particular to the parent $D$ meson
at rest, as in the case of the HPQCD simulation.  Moreover, it is argued in Ref.~\cite{Lubicz:2017syv} that the restricted kinematics employed in the simulations of  Refs.~\cite{Na:2010uf,Na:2011mc} 
may obscure the presence of hypercubic
effects in the lattice data,  and these corrections can affect the extrapolation to
the continuum limit in a way that depends on the specific lattice formulation. This might be one of the sources of the important discrepancies found between  the $D\to\pi$ form factors reported by the HPQCD and ETM Collaborations in the region close to $q_{\rm
max}^2=(m_D-M_\pi)^2$, as can be seen in the left top panel of Fig.~\ref{fig:combinedfit}.

Here, we prefer to fit to the most recent data together with the covariance
matrices provided by the ETM Collaboration. This analysis is based on  gauge configurations produced with $N_f=2+1+1$ flavors of
dynamical quarks at three different values of lattice
spacing, and with pion masses as small as 210 MeV. Lorentz symmetry breaking due to hypercubic effects is
clearly observed in the ETM data and included in the decomposition
of the current matrix elements in terms of additional form factors. Those discretization errors have not been considered in the HPQCD analyses, and for this reason we have decided to exclude the results of these latter collaboration in our fits. 

The scalar form factors involved in the $D\to \pi$ and $D\to \bar K$ transitions are related to the Omn\`es matrices displayed in Figs.~\ref{fig:out2cc} and \ref{fig:out3cc} through\footnote{Notice that the particle charges are not
specified in the notation used in Lattice QCD, for instance,
$D^0\to\pi^-$ in Eq.~(\ref{eq:mo.dpi}) is simplified to $D\to\pi$, to be used
below and denoted by $D\pi$ in the lattice paper.}  Eq.~\eqref{eq:FFmatrix} and Eqs.~\eqref{eq:mo.dpi} and \eqref{eq:mo.dk}. Hence, the bottom-charm combined $\chi^2$ now reads 
\bea
\chi^2&=&(\chi_{\rm cov}^2)^{\bar{B}\to\pi}_{\rm UKQCD}+(\chi_{\rm cov}^2)^{\bar{B}_s\to K}_{\rm UKQCD}+(\chi^2)^{\bar{B}\to\pi}_{\rm FL-MILC}
\nonumber\\
&+&(\chi^2)^{\bar{B}\to\pi}_{\rm HPQCD}+(\chi^2)^{\bar{B}_s\to K}_{\rm HPQCD}\nonumber\\
&+&(\chi^2)_{\rm LCSR}^{\bar{B}\to\pi} + (\chi^2)_{\rm LCSR}^{\bar{B}_s\to K} \nonumber \\
&+& (\chi_{\rm cov}^2)^{D\to\pi}+(\chi_{\rm cov}^2)^{D\to \bar{K}}
\ , \label{eq:chi2bottom-charm}
\eea
where we have added  sixteen ETM points, eight for each of the two $D\to \pi$ and $D\to \bar K$ decay modes. Each of the new eight-point sets is correlated   and the corresponding covariance matrices\footnote{The $D\to\pi$  and $D \to \bar K $ scalar form factor covariance-matrices have troublesome small
eigenvalues, as small as $10^{-6}$ or even $10^{-9}$. Due to this, the fitting procedure could be
easily spoiled since a tiny error in the fitting function yields a huge
$\chi^2$ value (specific examples can be found in Ref.~{\cite{Jang:2011fp}}). We have used 
the singular value decomposition (SVD) method to tackle this
issue, which is widely used by a number of lattice 
groups~\cite{Bailey:2010vz,Bhattacharya:1999uq,Bernard:2002pc}.}
 have been obtained from the authors of Ref.~\cite{Lubicz:2017syv}. Thus, we are fitting five parameters to a total of 43 points.  The best-fit results for the five unknown parameters and their Gaussian
correlation matrix are collected in Table~\ref{tab:fit.combined} and the resulting scalar form factors are
shown in Fig.~\ref{fig:combinedfit}. 

The results for the bottom scalar form factors are almost the same as the ones shown in Fig.~\ref{fig:bottomfit}, while the ETM $c \to d,s$ transition form factors are  remarkably well described within the present scheme. As in the former best-fit to only the $\bar B_{(s)}$ results, the large value obtained for $\chi^2/dof$ is mainly due to the existing tension   between the LQCD results from different collaborations in the $\bar B \to \pi$ decay.   

Due to the hypercubic
effects, there might be inconsistencies between the ETM and HPQCD analyses for the $D\pi$
scalar form factor in the region close to $q_{\rm max}^2=(m_D-M_\pi)^2$~\cite{Lubicz:2017syv}.  As one
can see, our result disagrees with the HPQCD data in that region too. We have
checked that if we fit to the HPQCD instead of the ETM data in the charm sector, the best fit still tends to coincide with the ETM data.
This observation is important and it seems to indicate that the Lorentz symmetry breaking effects
in a finite volume, due to the hypercubic artifacts, could be important in the
LQCD determination of the form factors in semileptonic heavy-to-light
decays, as pointed out in Ref.~\cite{Lubicz:2017syv}. 

\begin{figure*}[t]
\begin{center}
\makebox[0pt]{\includegraphics[width=0.45\textwidth]{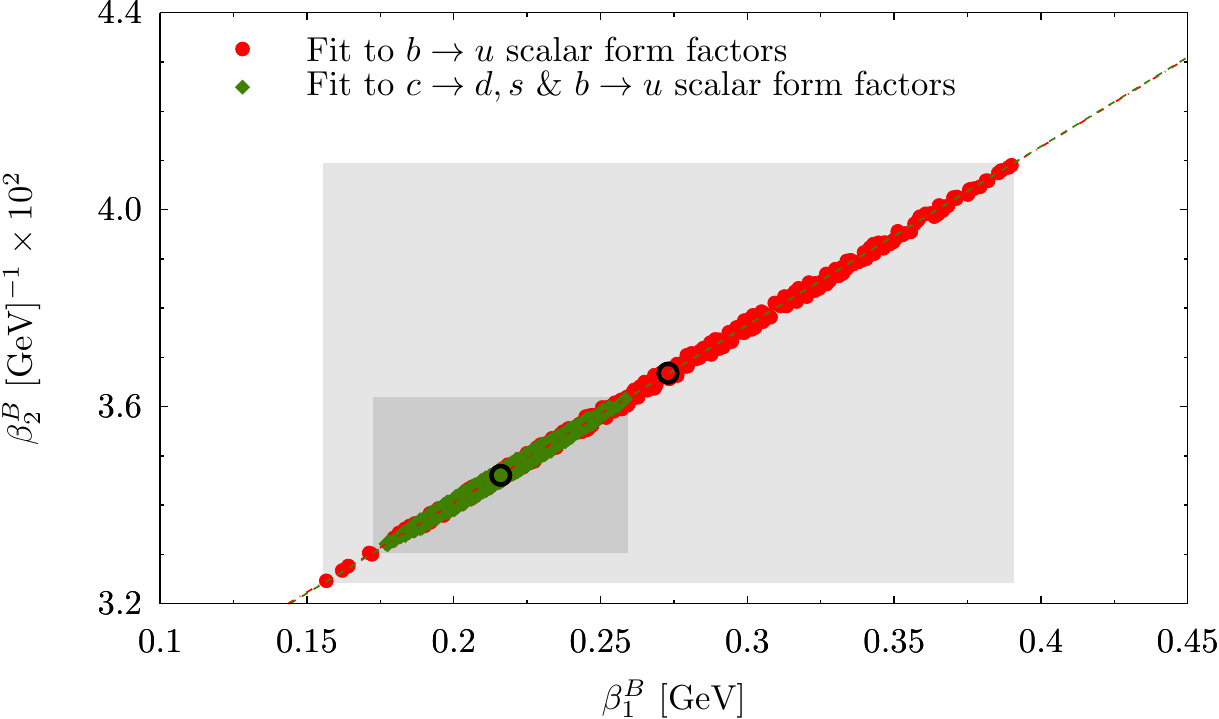}\hspace{0.5cm}\includegraphics[width=0.45\textwidth]{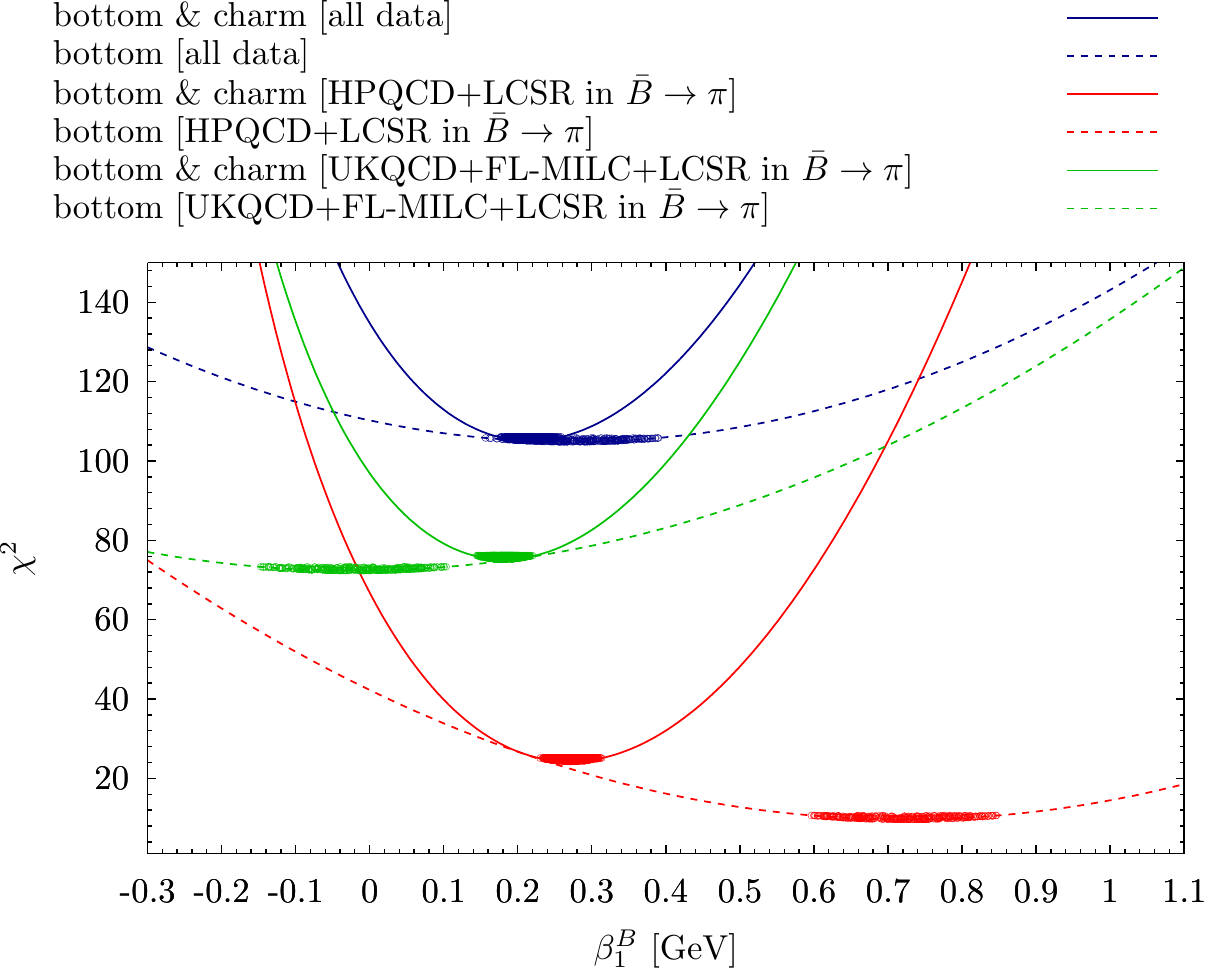}}
\caption{ Left plot: Statistical 68\% CL ellipses and one-sigma-rectangle bands in the $(\beta_1^B,\beta_2^B)-$plane obtained by minimizing  the merit functions given in Eqs.~\eqref{eq:chi2bottom} and \eqref{eq:chi2bottom-charm}. In addition the correlation relation of Eq.~\eqref{eq:recta} is also displayed. A similar relation deduced from the best-fit results of Table~\ref{tab:fit.combined} is also shown, though it is hardly distinguishable from the former one. Right plot: Dependence of $\chi^2$ on $\beta_1^B$ for different situations. The circles stand for the different best-fit results, accounting for variations of  $\chi^2$ up to one unit from the minimum value, while the dashed and solid curves have been obtained by relating $\beta_1^B$ and $\beta_2^B$ through Eq.~\eqref{eq:recta} and minimizing $\chi^2$ with respect to the other parameters, $\beta_0, \delta$ and $\delta^\prime$, respectively.  \label{fig:correlac}}
\end{center}
\end{figure*}

The HQFS breaking parameters $\delta$ and $\delta^\prime$ turn out to be quite correlated and their size is of the order $\Lambda_{\rm QCD}/m_c$.  As expected, $\delta$ presents also a high degree of correlation with $\beta_{1}^B$ and $\beta_2^B$, and  on the other hand, the combined fit does not reduce the large correlation between these two latter LECs, while the central values (errors) quoted for them in Table ~\ref{tab:fit.combined} are compatible within errors with (significantly  smaller than) those given in Eq.~\eqref{eq:b1b2besfit}, and obtained from the fit only to $b \to u$ transitions. In addition, the values quoted for $(\beta_{1}^B,\beta_2^B)$ in Table ~\ref{tab:fit.combined} perfectly lie in the straight line of Eq.~\eqref{eq:recta}, deduced from the fit to only bottom form factors carried out in the previous Subsect.~\ref{sec:fit}. 
Indeed, the straight line that one can construct with the results of Table ~\ref{tab:fit.combined} in the $(\beta_1^B, \beta_2^B)-$plane is practically indistinguishable from that  of Eq.~\eqref{eq:recta}. All this can be seen in the left plot of Fig.~\ref{fig:correlac}, where both straight lines are depicted, together with the statistical 68\% CL ellipses and the one-sigma-rectangle bands obtained by minimizing  the merit function given in Eq.~\eqref{eq:chi2bottom} or alternatively  in Eq.~\eqref{eq:chi2bottom-charm}, and considering only bottom or bottom and charm scalar form factors, respectively. 

In the right plot of Fig.~\ref{fig:correlac}, we show the dependence of $\chi^2$ on $\beta_1^B$ for different situations. We display the combined charm-bottom and the bottom-only fits, and in both cases, we have considered results obtained when all $\bar B \to \pi$ LQCD form factors  or only the HPQCD or the UKQCD and the FL-MILC subsets  of results are considered in the fits. The circles stand for the different best-fit results, accounting for variations of  $\chi^2$ up to one unit from the minimum value\footnote{Thus, the ranges marked by the circles show the statistical errors of $\beta_1^B$ in each fit.}, while the dashed and solid curves have been obtained by relating $\beta_1^B$ and $\beta_2^B$ through Eq.~\eqref{eq:recta} and minimizing $\chi^2$ with respect to the other parameters, $\beta_0, \delta$ and $\delta^\prime$, respectively. Several conclusions can be extracted from the results shown in the figure: 
\begin{itemize}
 
 \item The combined charm-bottom analyses (solid lines) provide large curvatures of  $\chi^2$ as a function of $\beta_1^B$, hence leading to better determinations of this latter LEC, always in the 0.2 GeV region, as we also found in Eq.~\eqref{eq:chi2bottom} from the best fit to only the bottom results. A value for $\beta_1^B$ close to this region, taking into account errors,   is  also found from a fit where only  the bottom form factors are considered, but without including  the HPQCD $\bar B \to \pi$  results (dashed-green curve). Only the  dashed-red line (fit only to the bottom results, but without including in this case the UKQCD and FL-MILC $\bar B \to \pi$ scalar form factors) turns out to be incompatible with the combined fit presented in Table~\ref{tab:fit.combined}. 
 Thus, we find some arguments to support the range of values quoted in Table~\ref{tab:fit.combined} for  the parameters $(\beta_1^B, \beta_2^B)$ that appear in the heavy meson chiral perturbation theory (HMChPT) expansion of the scalar form factors at the bottom scale.    
 \item The existing tension between HPQCD, and  UKQCD and FL-MILC sets of $\bar B \to \pi$ form factors leads to large values of $\chi^2$. Thus, as mentioned above, statistical errors should be taken with some care, and some systematic uncertainties would need to be considered in derived quantities, as for instance in the values of the form factors at $q^2=0$ or in the CKM mixing parameters. We note that this source of systematics also induces variations on the fitted parameters in Table~\ref{tab:fit.combined} which range between  50\% ($\beta_0$ and $\beta_2^B$) to 100\%   ($\beta_1^B$, $\delta$ and $\delta^\prime$) of the statistical errors quoted in the table. 
 \end{itemize}

Our predictions for the scalar form factors for the $D_s\to \eta$, $D\to \eta$ and $D_s\to K$
transitions, for which there are no lattice results as yet, are also shown in Fig.~\ref{fig:combinedfit}. Note that transitions involving the $\eta$ meson in the final state are more difficult to be evaluated in LQCD simulations.  Interestingly, the $D\to\eta$ scalar form factor in
the three-channel $(0,1/2)-$case is  largely suppressed, similar to the component regarding the $K\to\eta$ transition in the strangeness-changing scalar form factors as shown in Ref.~\cite{Jamin:2001zq}. 

 \begin{table}[t]
\caption{Results from the bottom-charm combined fit, with $\chi^2$ defined in Eq.~\eqref{eq:chi2bottom-charm} and a total of 38 degrees of freedom. The first set of errors in the best-fit parameters is obtained from the minimization procedure, assuming Gaussian statistics, while the second one accounts for the uncertainties of the LECs quoted in Ref.~\cite{Liu:2012zya} that enter in the definition of the chiral amplitudes. The LECs  $\beta_0$ and  $\beta_1^B$ ($\beta_2^B$) are given in units of GeV (GeV$^{-1}$).}\label{tab:fit.combined}
\begin{tabular}{cl|rrrrr}
\hline\hline
&  & \multicolumn{5}{c}{correlation matrix}\\
\multicolumn{2}{c|}{$\frac{\chi^2}{dof} =2.77$} & $\beta_0$ & $\beta_1^B$  & $\beta_2^B$ &$\delta$ & $\delta^\prime$\\
\hline
$\beta_0$           & $  0.152(14)(13)$                 & $ 1.000$ & $ 0.502$ & $ 0.499$ & $-0.490$ & $ 0.311$\\
$\beta_1^B$           & $  0.22(4)(4)$                  & $ 0.502$ & $ 1.000$ & $ 0.995$ & $-0.965$ & $ 0.848$\\
$\beta_2^B$           & $  0.0346(16)(15)$              & $ 0.499$ & $ 0.995$ & $ 1.000$ & $-0.958$ & $ 0.845$\\
$\delta$            & $  0.138(21)(18)$                 & $-0.490$ & $-0.965$ & $-0.958$ & $ 1.000$ & $-0.942$\\
$\delta^\prime$     & $ -0.18(4)(2)$                    & $ 0.311$ & $ 0.848$ & $ 0.845$ & $-0.942$ & $ 1.000$\\
\hline\hline
\end{tabular}
\end{table}


\subsubsection{Further considerations}

We have also obtained results using constant and quadratic MO polynomials. In the first case, the dispersive representations of the form factors should be matched to the LO chiral ones, where the terms driven by  the $\beta_1^P$ and $\beta_2^P$ LECs are dropped out. The first consequence is that bottom and charm sectors are no longer connected since, in addition,  we are not enforcing the heavy-quark scaling law for the decay constants. To better describe the data, one might perform separate fits to bottom and charm form factors with free $\vec{\alpha}_0$ parameters in Eq.~\eqref{eq:poly}. Fits obviously are poorer, and  moreover, they do not necessarily provide reliable estimates of the form factors at $q^2_{\rm max}$, since the fitted parameters are obtained after minimizing a merit function constructed out of data in the whole $q^2-$range accessible in the decays. For charm decays, the description of the 
$D\to \pi$ form factor is acceptable, while that of the $D\to \bar K$ is in comparison worse, mostly because the $s-$dependence induced by the $D_{s0}^\ast(2317)$ can not be now modulated by the MO polynomial. In the bottom sector, as one should expect, the simultaneous description of LQCD and LCSR form factors in the vicinity of $q^2_{\rm max}$ and $q^2=0$, respectively, becomes poorer. Indeed, since the LQCD input has a larger weight in the $\chi^2$ than the LCSR one, the latter form factors are totally missed by the new predictions, which now lie below the lower error bands of the LCSR results. 

The consideration of quadratic MO polynomials solves this problem, as shown in Fig.~\ref{fig:bottomfit-rank2} of Appendix~\ref{sec:quadMo}. Indeed, it is now possible to  improve the description of the  $\bar B\to \pi $ LCSR form factors, providing still similar results in the $q^2_{\rm max}-$region, where the LQCD data are available. Thus, for instance, we get  $f_+^{\bar{B}\to\pi}(0)=0.248 (10)$ using the new fit to be compared with $0.211 (10)$ obtained using the parameters of the fit of Eq.~\eqref{eq:b1b2besfit} (form factors displayed in Fig.~\ref{fig:bottomfit}). Nevertheless, as we will see in the next subsection, there exist some other systematic errors, which practically account for the latter difference, and thus this source of uncertainty will be considered in the determination of the CKM matrix element $|V_{ub}|$. In addition, though the $\chi^2/dof$  obtained with quadratic MO polynomials is better, it is still large (around 3.7) due to the  tension between the  $\bar B \to \pi$ LQCD 
results from different collaborations. Moreover $\beta_1^B$ and $\beta_2^B$ are still fully correlated, and the quadratic terms of the MO polynomials that multiply the elements $\Omega_{ij}$, $(i=1,2,3, ~j=2,3)$ of the matrix displayed in Fig.~\ref{fig:out3ccBott} are almost undetermined (see the large errors in the parameters $\alpha_{2,3}$ and especially $\alpha_{2,2}$ given in Table~\ref{tab:bottomfit-rank2}). 
Finally, the central-value predictions, that will be shown in Subsec.~\ref{sec:sfq2max},  for the form factors above  $q^2_{\rm max}$ and to moderate energies amenable to be described using the 
unitarized coupled-channel chiral approach, are not affected by the inclusion of quadratic terms in the MO polynomials, though errors are enhanced. For all of this, we  consider our best estimates for the form factors those obtained using rank-one polynomials.

We do not discuss quadratic terms in the charm-sector because rank-one MO polynomials  led  already to excellent reproductions of the form factors (see Fig.~\ref{fig:combinedfit}), 
in part due to the smaller $q^2-$range involved in these decays. Moreover a correct charm-bottom combined treatment will require the matching at 
next-to-next-to-leading order (NNLO) in the chiral expansion,  which is beyond the scope of this work.

\subsection{Extraction of CKM elements and predictions\label{sec:CKM}}

Taking advantage  that scalar and vector form factors are equal at $q^2=0$, the results of the combined charm-bottom fit presented in the previous subsection can be used to extract the vector form factor, $f_+$, at $q^2=0$ for various semileptonic decays studied in this work. Moreover, given some experimental input for the quantity $|V_{Qq}| f_+(0)$, with $Qq=bu,cd$ or $cs$, we can extract the corresponding CKM matrix element using the present MO scheme. Measurements of the differential distribution $d\Gamma(H\to\bar{\phi}\ell\bar{\nu}_\ell)/dq^2$ at $q^2=0$ will directly provide  model independent determinations of $|V_{Qq}| f_+(0)$,\footnote{Neglecting the lepton masses.} while measurements of the total decay width could  be used to estimate this latter quantity  only  after relying on some model for the $q^2-$dependence of $f_+$.

In the charm sector from the fit presented in Table~\ref{tab:fit.combined} and Fig.~\ref{fig:combinedfit}, we find
\begin{align}
f_+^{D\to\pi}(0)     & = 0.585(35)_{\rm stat}(19)_{\rm sys_1}(32)_{\rm sys_2},  \label{eq:charm-results1}\\
f_+^{D\to\bar{K}}(0) & = 0.765(30)_{\rm stat}(4)_{\rm sys_1}(14)_{\rm sys_2} \, , \label{eq:charm-results2}
\end{align}
where the first and second sets of errors are similar to those quoted in Table~\ref{tab:fit.combined} and account for
statistical (propagated from the $1\sigma$ fluctuations  of the fitted parameters) and chiral systematical (propagated from the errors of the  LECs that enter in the computation of the MO  matrix) uncertainties, respectively. The third set of errors (${\rm sys_2}$) takes into account the variations that are produced when in  the best fit one considers alternatively only  HPQCD or UKQCD and FL-MILC  $\bar B  \to \pi$ form factors. 
The results of Eq.~\eqref{eq:charm-results1} and  ~\eqref{eq:charm-results2} are in good agreement with our preliminary estimates reported in \cite{Yao:2017unm}, where we fitted only to the charm ETM LQCD form factors.

In combination with the experimental values 
\begin{align}
f_+^{D\to\pi}(0)|V_{cd}| & = 0.1426(19)~,\nonumber\\
 f_+^{D\to\bar{K}}(0)|V_{cs}| & = 0.7226(34)~,
\end{align}
taken from the report by the Heavy Flavor Averaging Group
(HFLAV)~\cite{Amhis:2016xyh}, we obtain 
\begin{align}
\label{eq:ckm1}
|V_{cd}| & = 0.244 (22)\\
|V_{cs}| & = 0.945 (41)
\end{align}
for the corresponding CKM matrix elements. The dominant error is the theoretical one affecting the determination of the form factors at $q^2=0$,  within the scheme presented in this work. As expected, these results nicely agree with those reported by the ETM Collaboration~\cite{Lubicz:2017syv} since we describe rather well the LQCD scalar form factors calculated in this latter work. The values of Eq.~\eqref{eq:ckm1} agree within around $1\sigma$ with the average ones\footnote{Determinations from leptonic and semileptonic decays, as well as from neutrino scattering data in the case of $|V_{cd}|$, are used to obtain the PDG averages.} given in Ref.~\cite{Olive:2016xmw},
\begin{align}
|V_{cd}| & = 0.220(5)~,\\
|V_{cs}| & = 0.995(16)~.
\end{align}
On the other hand, the test of the second-row unitarity of the CKM matrix is satisfied within errors
\bea
|V_{cd}|^2+|V_{cs}^2|+|V_{cb}|^2=0.95 (9)\ ,
\eea
where $|V_{cb}|=0.0405(15)$ from PDG~\cite{Olive:2016xmw} has been used.

Likewise, in the bottom sector, we obtain from the combined bottom-charm fit
\begin{equation}
f_+^{\bar{B}\to\pi}(0) = 0.208(7)_{\rm stat}(15)_{\rm sys_1}(30)_{\rm sys_2}\, , \label{eq:f+pi}
\end{equation}
where we see that the error budget is now  dominated by the inconsistency between HPQCD and UKQCD and FL-MILC sets of results for $f_0^{\bar{B}\to\pi}$ for high $q^2$, above 17 GeV$^2$. Dropping out the 
UKQCD and FL-MILC sets of results for $\bar{B}\to\pi$, the LCSR and HPQCD results for this transition can be significantly better described simultaneously, leading to values of the form factor at $q^2=0$ around 0.24 for the combined charm-bottom fit,  in the highest edge of the interval quoted in Eq.~\eqref{eq:f+pi},  and compatible within errors with the result of $0.26^{+0.04}_{-0.03}$ predicted in  Ref.~\cite{Duplancic:2008ix} using LCSR\footnote{Fitting only to the $b\to u$ data and not considering UKQCD and FL-MILC sets of $\bar{B}\to\pi$ results, we find $f_0^{\bar{B}\to\pi}(0)\sim$ 0.27, even in better agreement with the LCSR determination. }. However, the description of the $\bar B_s \to K$ and $D\to \pi$ scalar form factors gets somewhat worse, being thus the situation unclear. 

In principle, based on the above values, the CKM element $|V_{ub}|$ could be determined as in the charm sector.  However, the full kinematic region in the bottom case is very broad, and the experimental 
determination of $f_+(0)|V_{ub}|$ might suffer from large systematical uncertainties. A customary way to extract $|V_{ub}|$ has been to perform a joint fit to the LQCD and LCSR theoretical results for $f_+(q^2)$ and to measurements of the differential decay width, 
with $|V_{ub}|$ being a free parameter, see, e.g., Refs.~\cite{Ball:2006jz,Flynn:2006vr,Bourrely:2008za}. This is not feasible to us, since we only know the value of the vector form factor at zero momentum by using the relation $f_+(0)=f_0(0)$. However the latest Belle~\cite{Ha:2010rf} and BaBar~\cite{Lees:2012vv} works reported accurate measurements of the $\bar B \to \pi $  partial branching fractions in several bins of $q^2$ that are  used to extract the $f_+$ form factor shape and the overall normalization determined by $|V_{ub}|$. As a result, Belle and BaBar obtained  values of $(9.2 \pm 0.3)\times 10^{-4}$ and $(8.7 \pm 0.3)\times 10^{-4}$ for $f_+(0)|V_{ub}|$, respectively. Though the latter values were extracted from direct fits to data, they might be subject to some systematic uncertainties, since they were obtained using some specific $q^2$ parameterizations (Becirevic and Kaidalov~\cite{Becirevic:1999kt} and Boyd-Grinstein-Lebed~\cite{Boyd:1997qw} in the Belle and BaBar works, respectively). 
Nevertheless, we average both determinations and we take

\begin{equation}
f_+^{\bar{B}\to\pi}(0)|V_{ub}|=(8.9\pm0.3)\times 10^{-4}~,
\end{equation}
Using this latter value and our estimate for the form factor at $q^2=0$ given in Eq.~\eqref{eq:f+pi}, we get
\begin{align}
10^3|V_{ub}|&= 4.3 (7) \label{eq:vub}
\end{align}
There exist tensions between the inclusive and exclusive determinations of $|V_{ub}|$ \cite{Olive:2016xmw}:
\begin{align}
10^3|V_{ub}| & = 4.49(16)(17) & \text{(inclusive),}\\
10^3|V_{ub}| & = 3.72(19)               & \text{(exclusive).}
\end{align}
and combining both values, R.~Kowalewski and T.~Mannel quote an average value of 
\be
10^3|V_{ub}| = 4.09(39) 
\ee
in the PDG review~\cite{Olive:2016xmw}, which is in good agreement with our central $|V_{ub}|$ result of Eq.~\eqref{eq:vub}. We should mention that it is higher than the typical values obtained from  LQCD and LCSR determinations of the $\bar B \to \pi $ $f_+(q^2)$ form factor, combined with measurements of the $q^2$ distribution of the differential width.  Thus, the FLAG review~\cite{Aoki:2016frl} gives an average value (in $10^3$ units) of $3.67\pm 0.09 \pm 0.12$. Nevertheless, this latter value is still compatible, taking into account the  uncertainties, with our result.   

These extractions of the CKM elements rely strongly on the results either from LQCD in the high $q^2$ region or from LCSR in the vicinity of  $q^2=0$ (the latter only in the bottom sector), which are used in the combined fit, and hence are not {\it ab initio} predictions. However, our extractions incorporate the influence of general $S$-matrix properties,  in the sense that unitarity and analyticity are implemented in the MO representation of the scalar form factors. Moreover, one of the advantages of our approach is that we can make predictions for the channels related by chiral $SU(3)$ symmetry of light quarks. In some of these channels, the form factors  are difficult for LQCD due to the existence of disconnected diagrams of quark loops. The $D\to \eta$, $D_s\to K$, $D_s\to\eta$ and $\bar{B}\to\eta$ scalar form factors were already  shown in Fig.~\ref{fig:combinedfit} for the whole kinematical regions accessible in the decays.  On the other hand, their values at $q^2=0$  are particularly important, since 
they might serve as alternatives to determine the CKM elements when experimental measurements of the corresponding differential decay rates become available. Our predictions for the absolute values of the vector form factors at $q^2=0$ are (we remind once more here that vector and scalar form factors coincide at $q^2=0$)
\begin{align}
|f_+^{D\to\eta}(0)| & = 0.01(3)_{\rm stat}(2)_{\rm sys_1}(4)_{\rm sys_2},\\
|f_+^{D_s\to K}(0)| & = 0.50(6)_{\rm stat}(3)_{\rm sys_1}(5)_{\rm sys_2},\\
|f_+^{D_s\to\eta}(0)|& = 0.734(21)_{\rm stat}(21)_{\rm sys_1}(3)_{\rm sys_2},\\
|f_+^{\bar{B}\to\eta}(0)| & = 0.82(1)_{\rm stat}(7)_{\rm sys_1}(3)_{\rm sys_2},\\
|f_+^{\bar{B}_s\to K}(0)| & = 0.301(9)_{\rm stat}(11)_{\rm sys_1}(26)_{\rm sys_2}
\end{align}
For the decay $\bar{B}_s\to K$, we find, adding errors in quadrature, $f_+^{\bar{B}_s\to K}(0)=0.30 \pm 0.03$ in perfect agreement with the results obtained from the LCSR ($0.30^{+0.04}_{-0.03}$~\cite{Duplancic:2008tk}) and HPQCD ($0.32\pm 0.06$~\cite{Bouchard:2014ypa}) analyses, but about 1 sigma above the LQCD result of the UKQCD Collaboration~\cite{Flynn:2015mha}. The single-channel Omn\`es-improved constituent quark model study of Ref.~\cite{Albertus:2014gba} led to  $0.297 \pm 0.027$, which is also in good agreement with our result.

\subsection{Scalar form factors above the $q^2_{\rm max}-$region}
\label{sec:sfq2max}

\begin{figure*}[t]
\begin{center}
\makebox[0pt]{\includegraphics[width=0.89\textwidth]{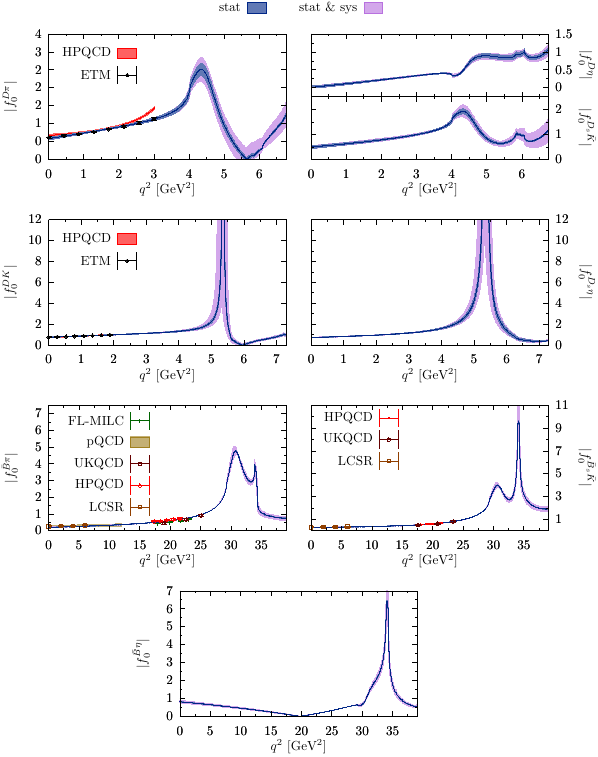}}
\caption{Scalar form factors for different $b\to u$ and $c\to d,s$ transitions. They have been  computed in this work using the MO matrices derived in Sec.~\ref{sec:MOsol} from the NLO HMChPT amplitudes  of    Refs.~\cite{Guo:2008gp, Liu:2012zya}, and the LECs, compiled in Table \ref{tab:fit.combined}, obtained from a fit to LQCD and LCSR results below $q^2_{\rm max}$.  Statistical (stat) and statistical plus systematic (stat \& sys)  68\%-confident level bands are also given and are calculated as explained in Fig.~\ref{fig:bottomfit}.}\label{fig:ffs-allrange}
\end{center}
\end{figure*}

It is worth recalling here the relation between the results obtained
for the form factors and the scattering amplitudes used as input of the MO representation. If we focus, for instance, on the charm form factors, the lightest open-charm
scalar resonance, called $D^\ast_0(2400)$ by the PDG~\cite{Olive:2016xmw}, lies in the  $(S,I)=(0,1/2)$ sector. In
Refs.~\cite{Kolomeitsev:2003ac,Guo:2006fu,Guo:2009ct}, two different states, instead of only one, were
claimed to exist in the energy region around the nominal mass of
the $D^\ast_0(2400)$. These studies were based on chiral
symmetry and unitarity. This complex structure should be reflected in the scattering regime of the form factors. Indeed, this can be seen in the first row of panels of Fig.~\ref{fig:ffs-allrange}, where form 
factors for different semi-leptonic transitions are shown above the $q^2_{\rm max}-$region. As discussed in Sec.~\ref{sec:MOsol}, here we use the $\mathcal{O}(p^2)$ HMChPT amplitudes  
obtained in Refs.~\cite{Guo:2008gp, Liu:2012zya}, which also successfully describe  the $(0,1/2)$ finite-volume energy levels reported in the recent LQCD simulation of Ref.~\cite{Moir:2016srx} (see Ref.~\cite{Albaladejo:2016lbb} for details) and are consistent with the precise LHCb data~\cite{Aaij:2016fma} for the angular moments of the $B^-\to D^+\pi^-\pi^-$~\cite{Du:2017zvv}. These chiral amplitudes predict the existence of two scalar broad resonances, 
instead of only one,  with masses around 2.1 and 2.45 GeV, respectively~\cite{Albaladejo:2016lbb, Du:2017zvv}, which produce some  signatures   in the $D\to \pi$, $D\to \eta$ and $D_s\to K$ form factors 
at around $q^2=4.4$ and 6 GeV$^2$, as can be appreciated in Fig.~\ref{fig:ffs-allrange}. The effect of this two-state structure is particularly visible in  the $D_s\to K$ form factor.  
Note that this two-state structure should have also some influence in the region below $q^2_{\rm max}$, where we have fitted the LQCD data. Below $q^2_{\rm max}$, the sensitivity of the 
form factors  to the  details of the two resonances is however smaller
than that of the energy levels calculated in the scattering region, since the
former ones are given below the lowest threshold, while the latter ones are
available at energies around and above it. Nonetheless, the success in describing the LQCD results for the $D\to \pi$ scalar form factor clearly supports  the chiral input, and the predictions deduced from it, used in the current scheme. If better
determined form factors were available in all of the channels, perhaps
the two state structure for the $D^\ast_0(2400)$ could be further and more accurately tested.

A similar pattern is found in the bottom sector~\cite{Albaladejo:2016lbb, Du:2017zvv}, as expected from the approximate heavy-flavor symmetry of QCD. The two-state structure is clearly visible, more than that in the charm sector, in the corresponding form factors (three bottom plots of Fig.~\ref{fig:ffs-allrange}), and it has a certain impact in the form factors close to $q^2_{\rm max}$, where LQCD results are available. 

In the charm $(S,I)=(1,0)$ sector the effect of the narrow $D_{s0}^\ast(2317)$ resonance, which is the $SU(3)$ flavor partner of the lighter one of the two $D^\ast_0$ states, predicted by the  unitarized NLO chiral amplitudes~\cite{Albaladejo:2016lbb, Du:2017zvv}, is clearly visible in the scalar $D\to \bar K$ and $D_s \to \eta$ form factors, and it fully dominates these form factors in the vicinity of the pole, as can be seen in the second row of panels of Fig.~\ref{fig:ffs-allrange}. Indeed, this state also influences the $D\to \bar K$ form factor below (near) $q^2_{\rm max}$, where the LQCD results are available\footnote{Indeed, the existence of the $D_{s0}^\ast(2317)$ was suggested in \cite{Flynn:2007ki} by fitting the single channel MO representation of the $D\to \bar K$ scalar form factor, constructed out the unitarized LO chiral elastic $DK$ amplitude,  to LQCD results of the scalar form factor below $q^2_{\rm max}$.}, and the excellent description of the ETM results gives clear support to the 
coupled-channel MO representation of the $D\to \bar K$  and $D_s\to \eta$ scalar form factors derived in this work.

\section{Summary and outlook\label{sec:CON}}

We have studied the scalar form factors that appear in  semileptonic heavy meson decays induced by the flavour-changing $b\to u$ and $c\to d, s$ transitions using the  MO formalism. The coupled-channel effects, due to re-scattering of the $H\phi$ ($H=D,\bar{B}$) system,  with definite strangeness and isospin,  are taken into account by solving coupled integral MO equations. We constrain the subtraction constants in the MO polynomials, which encodes the zeros of the form factors, thanks to light-quark chiral $SU(3)$ and heavy-flavor symmetries. 

The $H\phi$ interactions  used as input of  the MO equations are well determined in the chiral regime and are taken from previous work. In addition, some reasonably behaviors of the amplitudes at high energies are imposed, while   
appropriate heavy-flavor scaling rules are used to relate bottom and charm form factors. 
We fit our MO representation of the scalar form factors to the latest $c\to d , s$ and $b\to u$ LQCD and $b\to u$ LCSR results and determine all the involved parameters, in particular the two LECs ($\beta_1^P$ and $\beta_2^P$) that appear at NLO in the chiral expansion of the scalar and vector form factors near $q^2_{\rm max}$, which are determined in this work for first time.  We describe the LQCD and LCSR results rather well, and in combination with experimental results and using that $f_0(0)=f_+(0)$, we have also extracted the $|V_{ub}|$, $|V_{cd}|$ and $|V_{cs}|$ CKM elements, which turn out to be in good agreement with previous determinations from exclusive decays. 

We would like to stress that we describe extremely well the recent ETM $D\to \pi$ scalar form factor, which largely deviates from the previous determination by the HPQCD Collaboration,  providing an indication that the Lorentz symmetry breaking effects
in a finite volume, due to the hypercubic artifacts, could be important in the LQCD determination of the form factors in semileptonic heavy-to-light
decays, as claimed in Ref.~\cite{Lubicz:2017syv}. As it is also pointed out in the previous reference, this is a very
important issue, which requires further investigations, since it might become particularly relevant in the case of
the determination of the form factors governing semileptonic $\bar B-$meson decays into lighter mesons. 

We have also predicted the scalar form factors, which are in the same strangeness-isospin multiplets as the fitted $D\to\pi$, $D\to \bar K$, $\bar B \to \pi$ and $\bar B_s \to K$ ones. Our prediction of the form factors in such channels ($D\to \eta$, $D_s\to K$, $D_s\to\eta$, and $B\to\eta$) are difficult for LQCD simulations due to the existence of disconnected diagrams. These form factors are related to the differential decay rates of  different semileptonic heavy meson decays and hence provide alternatives to determine the CKM elements with the help of future experimental measurements. 

Moreover, we also find that the $D\to\eta$ scalar form factor is largely suppressed compared to the other two components ($D\to\pi$, $D\to \bar K$) in the three-channel $(0,1/2)-$multiplet, which is similar to what occurs for the $K\to\eta$ strangeness-changing scalar form factor in Ref.~\cite{Jamin:2001zq}. 

Our determination of the form factors has the advantage that the constraints from unitarity and analyticity of the $S$-matrix have been taken into account, as well as the state-of-the-art $H\phi$ chiral amplitudes. Thus, our predictions for the flavour-changing $b\to u$ and $c\to d, s$ scalar form factors above the $q^2-$region accessible in the  semileptonic decays, depicted in Fig.~\ref{fig:ffs-allrange}, should be quite accurate\footnote{Note that for the moderate $q^2-$values shown in Fig.~\ref{fig:ffs-allrange}, the form factors are largely insensitive to the high-energy input in the MO dispersion relation, and they are almost entirely dominated by the low-energy (chiral) amplitudes.} and constitute one of the most important findings of the current research. Indeed, we have shown how the form factors in this region reflect details of the chiral dynamics that govern the $H\phi$ amplitudes, and  that give rise to a new paradigm for heavy-light meson spectroscopy~\cite{Du:2017zvv} which questions the 
traditional $q\bar q$  constituent quark model interpretation, at least in the scalar sector.

As an outlook, the scheme presented here will also be useful to explore the $H\phi$ interactions by using the lattice data for the scalar form factors in semileptonic decays of $\bar{B}$ or $D$ mesons. As pointed out in Ref.~\cite{Du:2016tgp}, more data are needed to fix the LECs in the NNLO potentials. Since the dispersive calculation of $D\phi$ and $\bar{B}\phi$ scalar form factors depend on the scattering amplitudes of these systems, the LQCD results for the form factors can be used to mitigate the lack of data and help in the determination of the new unknown LECs. 

One might also try to extend the MO representation to a formalism in  a finite volume with unphysical quark masses, such that comparisons to the discretized lattice data could be directly undertaken. On the other hand, the chiral matching of the form factors can be carried out at higher order to take into account the expected sizable corrections in  $SU(3)$ HMChPT. Moreover, this improved matching will in practice suppose to perform additional subtractions in the dispersive representations of the form factors,  and it should reduce the importance of the high-energy input used for the $H\phi$ amplitudes. The high energy input turns out to be essential to describe the scalar $\bar B \to \pi$ form factor near $q^2=0$, and it represents one of the major limitations of the current approach.

Both  improvements would lead to a more precise and model-independent determination of the CKM matrix elements related to the heavy-to-light transitions.


\acknowledgments

DLY would like to thank Yun-Hua~Chen and Johanna~Daub for helpful discussions on
solving the MO problem. We would like to thank the authors
of Ref.~\cite{Lubicz:2017syv} for providing us the covariance matrices and J.~Gegelia for comments on the manuscript. 
P. F.-S. acknowledges financial support from the “Ayudas para contratos predoctorales para la formación de doctores” program (BES-2015-072049) from the Spanish MINECO and ESF.
This
research is supported by the Spanish Ministerio de Econom\'ia y Competitividad
and the European Regional Development Fund, under contracts FIS2014-51948-C2-1-P, FIS2017-84038-C2-1-P and SEV-2014-0398, by Generalitat Valenciana under contract
PROMETEOII/2014/0068, by the National Natural Science Foundation of China
(NSFC) under Grant No.~11747601, by NSFC and DFG though funds provided to the
Sino-German CRC 110 ``Symmetries and the Emergence of Structure in QCD'' (NSFC
Grant No. 11621131001),  by the Thousand Talents Plan for Young Professionals, by the
CAS Key Research Program of Frontier Sciences under Grant No.~QYZDB-SSW-SYS013, and by the CAS Center for Excellence in Particle Physics (CCEPP).

\appendix

\section{\boldmath Heavy-quark mass scaling of the LECs in the $D\phi$
interactions\label{sec:scaling.h}}

\begin{table*}[t!]
\caption{LECs and  subtraction constants used in this work to compute the UChPT $D\phi$  and  $\bar B\phi$ amplitudes\label{tab:lec}}
\begin{tabular}{cccccccc}
\hline\hline
  & $a(\mu=1~\text{GeV})$ & $h_0$ & $h_1$ & $h_{24}$ & $h_{4}^\prime$ & $h_{35}$ & $h_{5}^\prime$ \\
\hline
$D\phi$ & $-1.88^{+0.07}_{-0.09}$ & $0.014$ & $0.42$ & $-0.10^{+0.05}_{-0.06}$ & $-0.32^{+0.35}_{-0.34}$ & $0.25\pm0.13$ & $-1.88^{+0.63}_{-0.61}$  \\
$\bar B\phi$ & $-3.41^{+0.03}_{-0.04}$ & $0.038$ & $1.17$ & $-0.27\pm0.15$ & $-0.90^{+0.97}_{-0.93}$ & $0.68\pm0.36$ & $-5.23^{+1.74}_{-1.69}$  \\
\hline\hline
\end{tabular}
\end{table*}
Thanks to heavy quark symmetry, the $D\phi$ and $\bar{B}\phi$ interactions share the same effective Lagrangian with the correspondence $D\leftrightarrow \bar{B}$.  The NLO LECs $h_i$'s scale as~\cite{Liu:2012zya, Guo:2008gp, Cleven:2010aw}
\bea
h_{0,1,2,3}\sim m_Q\ ,\qquad h_{4,5}\sim \frac{1}{m_Q}\ ,
\eea
equivalently,
\bea
h_{0,1,2,3}^{B}=\frac{\bar{m}_B}{\bar{m}_D}h_{0,1,2,3}^{D}\ ,\qquad h_{4,5}^B=\frac{\bar{m}_D}{\bar{m}_B}h_{4,5}^D\ .
\eea
Here $\bar{m}_{D}$ ($\bar{m}_{B}$) is the average of the physical masses of the $D$ ($\bar{B}$) and  $D_s$ ($\bar{B}_s$) mesons. In addition, in the unitarized ChPT (UChPT) amplitudes there appears one  subtraction constant, $a(\mu)$, with $\mu=1$ GeV the scale introduced in dimensional regularization. In the $(S,I)=(0,1/2)$ channel, the subtraction constants in the charm, denoted by $a^D(\mu)$, and in the bottom, denoted by $a^B(\mu)$,  sectors are related as follows~\cite{Guo:2006fu,Albaladejo:2016lbb}:
\begin{enumerate}
\item First, given the phenomenological value of $a^D(\mu)$, a  sharp-cutoff, $q_{\rm max}$, is determined by requiring the dimensionally and the sharp-cutoff regularized $D_sK$ loop functions to be equal at threshold (see Eq.~(52) of Ref.~\cite{GarciaRecio:2010ki}). This cutoff turns out to be  $q_{\rm max}=0.72^{+0.07}_{-0.06}$ GeV. 
\item Next, $q_{\rm max}$ is used to determine $a^B(\mu)$ by requiring that the dimensionally and the sharp-cutoff regularized ${\bar B}_sK$ loop functions to be also equal at threshold. 
\end{enumerate}
LECs and subtraction constants for $D\phi$  and  $\bar B\phi$ interactions used in this work are collected in  Table~\ref{tab:lec}. Those in the charm sector are taken from Ref.~\cite{Liu:2012zya}.

\section{\boldmath Bottom form factors and quadratic MO polynomials  \label{sec:quadMo}}
\begin{figure*}[tbh]
\begin{center}
\includegraphics[width=0.9\textwidth]{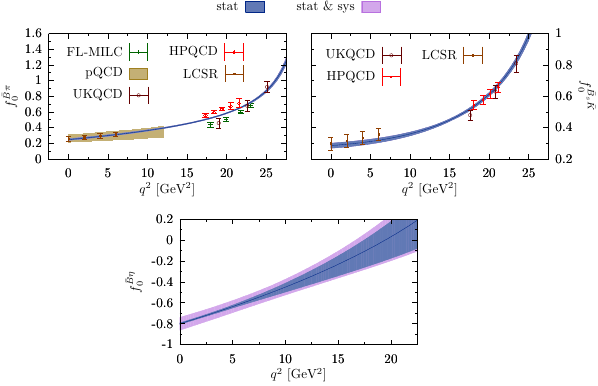}
\caption{Same as Fig.~\ref{fig:bottomfit}, but using a quadratic MO polynomial (see Eq.~\eqref{eq:poly2}), and parameters given in Table~\ref{tab:bottomfit-rank2}.\label{fig:bottomfit-rank2} }
\end{center}
\end{figure*}

 \begin{table}[t]
\caption{Results from  the fit to the scalar $\bar B \to \pi$  and $\bar B_s \to K$ LQCD \& LCSR form factors  using rank-two MO polynomials (see Eq.~\eqref{eq:poly2}), and  with $\chi^2$ defined in Eq.~\eqref{eq:chi2bottom}. There is a  total of 22 degrees of freedom and the best-fit gives $\chi^2/dof= 3.7$. The LECs  $\beta_1^B$, $\beta_2^B$ and $\alpha_{2,i}$ are given in units of GeV, GeV$^{-1}$ and GeV$^{-4}$, respectively.}\label{tab:bottomfit-rank2}
\begin{tabular}{ccccc}
\hline\hline
$\beta_1^B$  & $(\beta_2^B \times 10)$ &$(\alpha_{2,1}\times 10^{3})$ & $(\alpha_{2,2}\times 10^{3})$  &  $(\alpha_{2,3}\times 10^{3})$\\
\hline
0.74(22)     & 0.53(8) & 0.24(6) & $-0.1(7)$ & 1.0(8)\\
\hline\hline
\end{tabular}
\end{table}
We show in Fig.~\ref{fig:bottomfit-rank2}, the scalar $\bar B \to \pi,\eta$  and $\bar B_s \to K$ form factors obtained using rank-two MO polynomials. Specifically, we replace Eq.~\eqref{eq:poly} by 
\bea\label{eq:poly2}
\vec{\mathcal{P}}(s)=\vec{\alpha}_0+\vec{\alpha}_1\,s +\vec{\alpha}_2\,s^2\ ,
\eea
where $\vec{\alpha}_2$, together with $\beta_{1,2}^B$, are fitted to the merit function defined in Eq.~\eqref{eq:chi2bottom}. (Note that $\vec{\alpha}_{0,1}$ are still expressed in terms of $\beta_{1,2}^B$.)
Best fit results are compiled in Table~\ref{tab:bottomfit-rank2}. 
\bibliography{Dl3bib}

\end{document}